\documentclass[aps,prb,nofootinbib,preprintnumbers,notitlepage,twocolumn]{revtex4-2}

\usepackage{amsmath,amssymb,amsfonts,graphicx}
\usepackage{color}
\usepackage{subfigure}
\usepackage{bbm}
\usepackage[hidelinks]{hyperref}

\renewcommand{\vec}[1]{{\boldsymbol{#1}}}
\newcommand{\beq}{\begin{eqnarray}}
\newcommand{\eeq}{\end{eqnarray}}
\newcommand{\tabi}{\hspace{.1\textwidth}}

\begin{document}

\title{Large enhancement of Edelstein effect in Weyl semimetals \\ from Fermi-arc surface states}
\author{Kridsanaphong Limtragool}\thanks{Corresponding author}\email{kridsanaphong.l@msu.ac.th}
\author{Krisakron Pasanai}
\affiliation{Theoretical Condensed Matter Physics Research Unit, Department of Physics, Faculty of Science, Mahasarakham University, Khamriang Sub-District, Kantharawichai District, Maha-Sarakham 44150, Thailand}
%\date{\today}

\begin{abstract}
	
One remarkable feature of Weyl semimetals is the manifestation of their topological nature in the form of the Fermi-arc surface states. In a recent calculation by \cite{Johansson2018}, the current-induced spin polarization or Edelstein effect has been predicted, within the semiclassical Boltzmann theory, to be strongly amplified in a Weyl semimetal TaAs due to the existence of the Fermi arcs. Motivated by this result, we calculate the Edelstein response of an effective model for an inversion-symmetry-breaking Weyl semimetal in the presence of an interface using linear response theory. The scatterings from scalar impurities are included and the vertex corrections are computed within the self-consistent ladder approximation. At chemical potentials close to the Weyl points, we find the surface states have a much stronger response near the interface than the bulk states by about one to two orders of magnitude. At higher chemical potentials, the surface states' response near the interface decreases to be about the same order of magnitude as the bulk states' response. We attribute this phenomenon to the decoupling between the Fermi arc states and bulk states at energies close to the Weyl points. The surface states which are effectively dispersing like a one-dimensional chiral fermion become nearly nondissipative. This leads to a large surface vertex correction and, hence, a strong enhancement of the surface states' Edelstein response.

\end{abstract}
\maketitle

\section{Introduction}

Weyl semimetals\cite{Wan2011,Burkov2011,Weng2015,Huang2015,Lv2015,Xu2015,Armitage2018} have recently attracted a great attention because they provide a solid-state realization of Weyl fermions\cite{Weyl1929}. Weyl semimetals exhibit many unconventional properties that deviate from the standard theories of metals and semiconductors. For example, the chiral anomaly, the nonconservation of the chiral charge, leads to anomalous transport properties such as the chiral magnetic effect \cite{Zyuzin2012,Chernodub2014} and the negative magnetoresistivity \cite{Son2013,Burkov2015,Li2017}. Weyl fermions with broken time-reversal symmetry show the anomalous Hall effect which has a universal form that depends on the distance between Weyl nodes\cite{Burkov2014, Haldane2014}.

One of the manifestations of the topological properties in Weyl semimetals is the appearance of the nontrivial Fermi-arc surface states \cite{Wan2011, Haldane2014}. In the semimetallic phase, the Fermi arc is an open curve that terminates at the two Weyl nodes with opposite chiralities. When the chemical potential moves away from the Weyl points, the Fermi arc still survives and connects the two disjointed bulk Fermi surfaces enclosing the two Weyl nodes. These Fermi arcs on the surface were observed experimentally in TaAs with angle-resolved photoemission spectroscopy (ARPES) \cite{Lv2015,Xu2015}. An effective two-band model that describes a Weyl-node pair and the generation of the Fermi arc was proposed by Ref. \cite{Okugawa2014}. The Bloch Hamiltonian of the model is given by
\begin{align} \label{eq:weyl_okugawa_hamiltonian}
	H(\vec k) = \gamma(k_{x}^2 - m)\sigma_x  + v (k_{y} \sigma_y + k_{z}\sigma_z),
\end{align}
where $\sigma_i$ are the Pauli matrices acting on pseudospins, and $m$, $\gamma$, and $v$ are constant. In this model, the system is a Weyl semimetal if $m > 0$ and a band insulator if $m<0$. When a surface is introduced, Ref. \cite{Okugawa2014} showed that there is a Fermi-arc state on the surface in the phase $m>0$. 

Given that surfaces of a material are always present in a device application, it is crucial to understand the interplay between bulk states and the Fermi-arc surface states and how they affect the properties of Weyl semimetals. There are some studies that focused on the effect of Fermi arcs on the electrical transport. Ref. \cite{Gorbar2016} investigated the electrical conductivity of the surface states, $\sigma_s$, in the present of a quenched disorder using Kubo formalism. Since the Fermi arc states can be effectively described by a one-dimensional chiral fermion, one expects that their electrical transport should be dissipationless. However, from the calculation by \cite{Gorbar2016}, $\sigma_s$ is finite. $\sigma_s$ is maximum at the surface and decreases as one moves further away from the surface. Ref. \cite{Gorbar2016} explained this phenomenon with the existence of the gapless bulk states. The surface and bulk states are still coupled and so there are scatterings from the surface states to the bulk states. This results in a dissipative transport of the surface states. A study by Ref. \cite{Breithkreiz2019} tried to understand the contribution of Fermi arcs to the total electrical conductivity (i.e., the sum of bulk and surface conductivities) of a time-reversal-invariant Weyl semimetal in a finite-size geometry. Ref. \cite{Breithkreiz2019} split the system into the sum of a subsystem with broken time-reversal symmetry plus its time-reversal conjugate. Using the Landauer-type approach, they found that surface states' electrical conductivity could be as large as the bulk conductivity. These studies highlight the significant effects that the Fermi arcs have on transport properties of Weyl semimetals. 

The focus of this paper is the current-induced spin polarization or the Edelstein effect\cite{Edelstein1990}. A system with a strong spin-orbit coupling, such as a Rashba system and a topological insulator, for which the spin degeneracy is lifted, is expected to exhibit this effect\cite{Johansson2016}. In such a system, an electric field can be used to induced a perpendicular spin-polarization or magnetization inside a material. This effect could potentially be useful for applications in spintronics because it allows a manipulation of a magnetization with an electric field inside a nonmagnetic material. Using the semiclassical Boltzmann theory, Ref. \cite{Johansson2018} computed an electric-field-induced magnetic moment in a Weyl semimetal TaAs. They found the surface Edelstein effect of Weyl semimetals to be much stronger than Rashba systems and topological insulators. Furthermore, the magnetic moment of the surface states near the surface was found to be greater than that of the bulk states by about two orders of magnitude. This large surface Edelstein effect was attributed to long momentum relaxation times of the surface states. Additionally, an experiment performed on a Weyl semimetal WTe$_2$ indicated that the material exhibits a large charge-to-spin conversion effect\cite{Zhao2020}.

In this work, we further investigate the behavior in which the surface Edelstein response is much stronger than the bulk response near an interface with Kubo formalism. We calculate the magnetoelectric susceptibility of a Weyl semimetal with broken inversion symmetry. The perturbation theory techniques from \cite{Gorbar2016} are used to compute the surface self-energy and susceptibility. We include short-range scalar impurities in the model and compute vertex corrections within the self-consistent ladder approximation. In order to make a comparison with the surface's result, we also compute the bulk states' susceptibility. We show that, near an interface, the surface states have a much stronger Edelstein response than the bulk states when the chemical potentials are close to the Weyl points. At high chemical potentials, the Edelstein response of the surface states is about the same order of magnitude as that of the bulk states. At work here is the decoupling between the bulk and surface states close to the energy of the Weyl nodes. This can be seen from our calculation that the rate in which the surface states scatter into the bulk states vanishes at $\omega = 0$. The Fermi-arc states, effectively a chiral Fermion in 1D, must be almost dissipationless at low chemical potentials. This results in a large enhancement of the surface states' vertex correction and Edelstein response.

\section{Model of a Weyl semimetal with Broken Inversion Symmetry}

In this paper, we consider the model that Ref. \cite{Johansson2018} used to describe Weyl semimetals with broken inversion symmetry. Eq. \ref{eq:weyl_okugawa_hamiltonian} is modified by introducing two pseudospin sectors denoted with $p = \pm1$. Each sector contains a pair of Weyl nodes centered at $p\vec k_0$. Unlike Eq. \ref{eq:weyl_okugawa_hamiltonian}, the Pauli matrices $\sigma_i$ act on spins, but not on pseudospins. The Bloch Hamiltonian of this model is given by
\begin{align} \label{eq:weyl_hamiltonian}
	H_p(\vec k) = p\gamma(k_{p,x}^2 - m)\sigma_y  - v_y k_{p,y} \sigma_x + v_z k_{p,z}\sigma_z,
\end{align}
where $m$, $\gamma$, $v_y$, and $v_z$ are positive constants.  Here, $\vec k_{p}$ is a momentum with respect to $p \vec k_0$ (i.e., $\vec k_{p} = \vec k - p\vec k_0$). For simplicity of the notation, the subscript $p$ will be dropped in subsequent mentions of $\vec k_p$. One can show that this model is time-reversal invariant, but not invariant under inversion. Under the global time-reversal operation, the Hamiltonians of the two pseudospin sectors exchange. Since momenta and spins are coupled, this model is expected to show the Edelstein effect. Furthermore, as stated in \cite{Johansson2018}, such couplings can generate the spin polarization and texture which are the features observed experimentally in TaAs \cite{Xu2016}. We note that Ref. \cite{Johansson2018} applied the Hamiltonian of a form given in Eq. \ref{eq:weyl_hamiltonian} to TaAs which has 24 Weyl nodes in the Brillouin zone. They did not explicit include the couplings between different Weyl-node pairs in the Hamiltonian, but considered the scatterings due to impurities among the pairs. In this paper, we consider a simpler situation, i.e., our model only consists of 4 Weyl nodes (two nodes for each pseudospin sector $p$) from Eq. \ref{eq:weyl_hamiltonian}.

\begin{figure}
	\centering
	\includegraphics[scale=0.22]{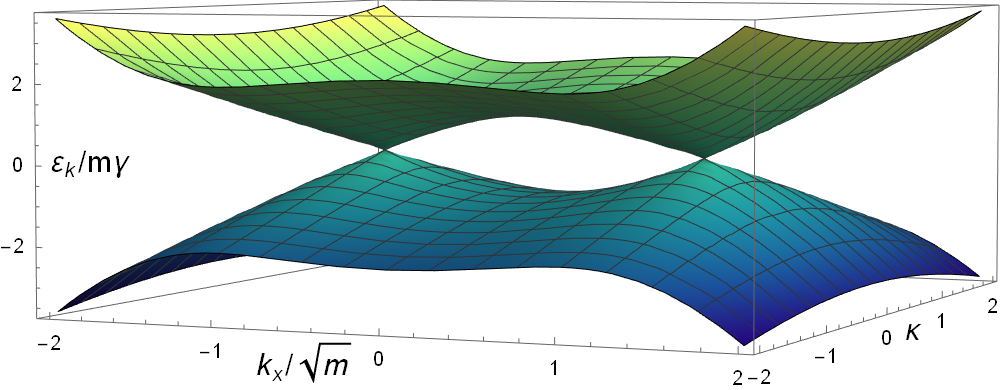}
	\caption{Bulk-state dispersion of a pair of Weyl nodes (Eq. \ref{eq:eigenenergy}). One of the axes in this plot is defined by $\kappa \equiv \frac{\sqrt{(v_yk_y)^2 + (v_zk_z)^2}}{m\gamma}$.} \label{fig:weyldis}
\end{figure}

The eigenenergy of this model,
\begin{align} \label{eq:eigenenergy}
		\varepsilon_k = \pm \sqrt{\gamma^2(k^2_x - m)^2 + v_y^2k_y^2 + v_z^2 k_z^2},
\end{align}
is displayed in Fig. \ref{fig:weyldis}. There are two Weyl nodes located at $(k_x,k_y,k_z) = (\pm \sqrt{m},0,0)$. When the chemical potential $\mu$ is small, there are two distinct closed bulk Fermi surfaces enclosing each individual node. As $\mu$ increases to $m\gamma$, the system undergoes a Lifshitz transition, at which point the two Fermi surfaces coalesce into one closed surface.

\section{Weyl semimetal/vacuum interface} \label{sec:surface}
To study the effects of the surface, let us assume that the system is a Weyl semimetal (as described by Eq. \ref{eq:weyl_hamiltonian}) in the region $z>0$ and a vacuum for $z<0$. This setup can be achieved by allowing $m$ to depend on $z$,
\begin{align}
m(z) = 
\begin{cases}
	m > 0, & z>0 \\
	-\tilde{m} \rightarrow -\infty, & z<0.
\end{cases}
\end{align}
Under this assumption, the surface of this system, which is a plane located at $z = 0$, breaks the translational invariance along the $z$ direction and, hence, the momentum along the $z$ direction is no longer conserved. $k_{z}$ in Eq. \ref{eq:weyl_hamiltonian} is replaced by $-i \frac{\partial}{\partial z}$. As a result, the time-independent Schr\"odinger equation of this system is given by
\begin{align} \label{eq:weyl_schrodinger}
\left( p\gamma(k_{x}^2 - m(z))\sigma_y   - v_y k_{y} \sigma_x -i v_z \sigma_z\frac{\partial}{\partial z} \right) \psi = \varepsilon \psi
\end{align}
Depending on the boundary conditions, this equation can be solved to yield the surface or bulk eigenstates.

\subsection{Surface states} \label{sec:surface_state}
The boundary conditions for the case of the surface states are the continuity of the wave functions at the surface, and the vanishing of the wave functions infinitely far away from the surface (see Eqs. \ref{eq:surfacebc1} and \ref{eq:surfacebc2} in Appendix \ref{app:hermitian_bc}). The normalization condition is
\begin{align}
\int^{L/2}_{-L/2} dx \int^{L/2}_{-L/2} dy \int_{0}^{\infty} dz \ \psi^\dagger(\vec r)\psi(\vec r)
= 1,
\end{align}
where $L \rightarrow \infty$ is the size of Weyl semimetals along the $x$ and $y$ directions. One can solve Eq. \ref{eq:weyl_schrodinger} (see Appendix \ref{app:surface}) to find the eigenenergy,
\begin{align} \label{eq:surface_states_energy}
\varepsilon^s_k = p v_y k_{y},
\end{align}
and the corresponding eigenstates,
\begin{align} \label{eq:eigenstates_surface_final}
\psi_{s}(\vec r) = 
\frac{\sqrt{\ell(k_x)}}{L}
\begin{pmatrix} 
	-p \\ 1
\end{pmatrix} e^{-\ell(k_x) z + ik_{x}x + ik_{y}y},
\end{align}
where $z > 0$ and $\sqrt{m} < k_x < \sqrt{m}$. Here, we define $\ell(k_x) \equiv \frac{\gamma}{v_x}(m - k_x^2)$ to simplify the above expression. Since $\ell(k_x) > 0$, $\psi_s(r)$ is exponentially decayed as one expects for a surface wave function. For a given value of chemical potential, the projection of the surface states' Fermi surface onto the plane $z = 0$ is
 a line which touches the two bulk Fermi surfaces. Such a projection of a fermi surface is known as the Fermi arc.

\subsection{Bulk states in the presence of a surface}

In the case of the bulk states, their wave functions are extended throughout the materials. The wave functions do not vanish as $z \rightarrow \infty$ unlike the case of the surface states. In order for the Hamiltonian to be hermitian, one requires that the ratio of the two components of the wave functions is a complex phase at $z = \infty$ (see Appendix \ref{app:hermitian_bc}). We choose this complex phase $e^{i\phi} = p$ (the pseudospin index) and interpret $z \rightarrow \infty$ as $z = L_z$, where $L_z$ can be thought of as the thickness of the Weyl semimetal slab.  Hence, the boundary condition at $z = L_z$ is $\psi_1(L_z)/\psi_2(L_z) = p$.

The normalization condition in the case of the bulk eigenstates is
\begin{align}
\int^{L/2}_{-L/2} dx \int^{L/2}_{-L/2} dy \int_{0}^{L_z} dz \psi_b^\dagger(\vec r) \psi_b(\vec r) = 1.
\end{align}
Solving Eq. \ref{eq:weyl_schrodinger} in the bulk case (see Appendix \ref{app:bulk}), one finds that the eigenenergy still has the same form,
\begin{align}
\varepsilon^b_k = \pm\sqrt{(v_z \ell(k_x))^2 + (v_y k_{y})^2 + ( v_z \tilde{k}_z)^2},
\end{align}
as Eq. \ref{eq:eigenenergy}. The difference is that the momentum along the $z$ direction is replaced by a discrete and positive quantum number $\tilde{k}_z$.\footnote{One can choose $\tilde{k}_z$ to be either strictly positive or strictly negative. For definiteness, we choose $\tilde{k}_z$ to be positive in this paper.} In the limit $L_z \rightarrow \infty$, $\tilde{k}_z$ becomes continuous. The corresponding bulk eigenstates in the presence of the surface located at $z = 0$ are
\begin{align} \label{eq:psi_b}
\psi_b(\vec r) =
A(\vec k_{\|},\tilde{k}_z)  e^{i(\vec k_{\|}\cdot\vec r_{\|} + \tilde{k}_z z)}  - A(\vec k_{\|},-\tilde{k}_z) e^{i(\vec k_{\|}\cdot\vec r_{\|} - \tilde{k}_z z)}, \nonumber \\
\end{align}
where 
\begin{align}
A(\vec k_{\|},\tilde{k}_z) \equiv & \frac{ - v_y k_{y} + i  pv_z \ell(k_x) +p\varepsilon + p v_z \tilde{k}_z }{\sqrt{8V_w \varepsilon (\varepsilon - p v_y k_{y})(\varepsilon^2 - (v_z \tilde{k}_z)^2)}} \nonumber \\
& \times\begin{pmatrix}
	v_y k_{y} - i  pv_z\ell(k_x) \\
	-\varepsilon +  v_z \tilde{k}_z
\end{pmatrix}.
\end{align}
Here, $\vec k_{\|} \equiv (k_x, k_y)$ and $\vec r_{\|} \equiv (x,y)$ are the momentum and position vectors parallel to the surface, respectively. $V_w \equiv L^2 L_z$ is the volume of the Weyl semimetals. Eq. \ref{eq:psi_b} can be understood as the superposition of the plane waves that are incident on and reflecting from the surface. Such an addition of the plane waves is required in order for the wave functions to satisfy the boundary conditions.

\section{Green function} \label{sec:green_function}
The Green function can be computed from eigenstate wave functions, $\psi_n(\vec r)$, and eigenenergy, $E_n$, by
\begin{align} \label{eq:green_func_eigenstates}
G(\omega,\vec r, \vec r') = \sum_{n}\frac{\psi_n(\vec r) \psi_n^\dagger (\vec r')}{\omega - E_n}
\end{align}
with the sum being over all eigenstates of the Hamiltonian. The summation here can be separated into the sum over bulk and surface eigenstates. This means the total Green function is
\begin{align}
G(\omega,\vec r, \vec r') = G_s(\omega,\vec r, \vec r') + G_b(\omega,\vec r, \vec r'),
\end{align}
where $G_s$ and $G_b$ are the surface and bulk Green functions, respectively.

For the case of the surface Green function, we substitute the surface eigenstates from Eq. \ref{eq:eigenstates_surface_final} into Eq. \ref{eq:green_func_eigenstates}. The summation over $E$ turns into an integral over $k_{x}$ and $k_{y}$. Performing Fourier transform and rewriting the matrix in the form of the identity and Pauli matrices, one finds the expression for the surface Green function is
\begin{align} \label{eq:green_func_surface}
G_s^{0}(\omega,\vec k_\parallel,z,z') = \ell(k_x)\frac{\mathbbm{1} - p\sigma_x}{\omega - p v_y k_{y}}e^{-{\ell(k_x)} (z+z')}.
\end{align}

We calculate the bulk Green function by substituting Eq. \ref{eq:psi_b} into Eq. \ref{eq:green_func_eigenstates} and then simplifying the results (see Appendix \ref{app:bulk}). We find that the bulk Green function has a form
\begin{align} \label{eq:Gb_Gb0_Gb1}
G_{b}^0(\omega,\vec k_{\parallel},z,z') = G_{ti}^0(\omega,\vec k_{\parallel},z,z')+G_{ni}^0(\omega,\vec k_{\parallel},z,z'), 
\end{align}
where
\begin{align} \label{eq:Gb0}
G_{ti}^0(\omega,\vec k_{\parallel},&z,z') = \int_{-\infty}^{\infty} \frac{d\tilde{k}_z}{2\pi} e^{i\tilde{k}_z(z - z')} \nonumber \\
&\times \frac{ \omega \mathbbm{1} +  v_z \tilde{k}_z \sigma_z -  v_y k_{y} \sigma_x - pv_z\ell(k_x)\sigma_y }{\omega^2 - \varepsilon_k^2}
\end{align}
and
\begin{align} \label{eq:Gb1}
G_{ni}^0(\omega,&\vec k_{\parallel},z,z') = \int_{-\infty}^{\infty} \frac{d\tilde{k}_z}{2\pi} e^{i\tilde{k}_z(z + z')}\frac{1}{(\omega^2 - \varepsilon_k^2)} \nonumber \\ &\times \Bigg[ \frac{p v_y \tilde{k}_z k_{y} - i\ell(k_x)\omega}{\tilde{k}_z - i\ell(k_x)}\mathbbm{1} - \frac{p \tilde{k}_z \omega + i v_y k_{y}\ell(k_x)}{ \tilde{k}_z - i\ell(k_x)}\sigma_x \nonumber \\ & \ \ \ \ \ \   - ipv_z(\tilde{k}_z + i\ell(k_x))\sigma_y \Bigg].
\end{align}
The term $G_{ti}^0$ is invariant under translation because it depends on the difference between $z$ and $z'$. $G_{ti}^0$ can be further Fourier transformed into a form
\begin{align} \label{eq:Gb0_2}
G_{ti}^0(\omega,\vec k) = \frac{\omega \mathbbm{1} +  v_z \tilde{k}_z \sigma_z -  v_y k_{y} \sigma_x - pv_z\ell(k_x)\sigma_y}{\omega^2 - \varepsilon_k^2} 
\end{align}
which is precisely the bulk Green function of Weyl semimetals without the interface. The correlation between two eigenstate wave functions which propagate in the same direction (i.e. between two incident waves or between two reflecting waves) gives rise to this translationally symmetric portion of the bulk Green function. On the other hands, $G_{ni}^0$ is not invariant under translation since it depends on the sum of $z$ and $z'$. This part of the Green function originates from a correlation between the incident and reflecting waves on the surface. In the absence of the surface, such a correlation would be zero. We note that an alternative method to calculate the full Green function of Weyl semimetals occupying half of the three-dimensional space was reported in \cite{Faraei2018}. Ref. \cite{Faraei2018} solved the differential equation with generalized hard-wall boundary conditions and found that the total Green function has a similar form to our result in this section, i.e., the sum of a translationally invariant term and a nontranslationally invariant term.

\section{Surface states' scattering rate from random impurity scattering} \label{sec:surface_selfen}

As in \cite{Johansson2018}, we study the Edelstein effect of Weyl semimetals in the presence of short-range random impurities. We use the same quench disorder model as Ref. \cite{Gorbar2016}. The impurities are assumed to be diluted, so that the perturbation theory we use to calculate self-energies, vertex corrections, and response functions are valid. In this model, electrons interact with the scalar impurities through the Hamiltonian,
\begin{align} \label{eq:Himp}
H_{\mathrm{I}} = \int d^3r \psi^\dagger(\vec r) U(\vec r)\psi(\vec r),
\end{align}
where $U(\vec r) = \sum_{i}u(\vec r - \vec r_i)$ is the potential of all impurities in the system and $u(\vec r - \vec r_i) = u_0 \delta(\vec r - \vec r_i)$ is a potential of an individual impurity centered at $\vec r_i$. Here, $u_0$ is a parameter with units of energy multiplied by volume. These impurities are assumed to be uniformly distributed in the sample. The quantities calculated from the impurity potentials such as a Green function depend on the positions of all impurities. In order to extract a meaningful result, one performs a disordered average by
\begin{align} \label{eq:dis_avg}
	\langle A \rangle_{\mathrm{dis}} = \int \prod^{N}_{i=1} d^3r'_i \rho(\vec r'_{1}, \vec r'_{2}, \vec r'_{3}, ..., \vec r'_{N})A(\vec r'_{1}, \vec r'_{2}, \vec r'_{3}, ..., \vec r'_{N}). \nonumber \\
\end{align}
where $A(\vec r'_{1}, \vec r'_{2}, \vec r'_{3}, ..., \vec r'_{N})$ is a function that depends on $N$ impurity positions and $\rho$ is the ensemble distribution function. From the assumption that the impurities are uniformly distributed, $\rho$ is simply
\begin{align}
\rho(\vec r'_{1}, \vec r'_{2}, \vec r'_{3}, ..., \vec r'_{N}) = \prod^N_{i=1} \frac{1}{V} = \frac{1}{V^N}.
\end{align}

\begin{figure}
	\centering
	\subfigure[\label{fig:gug}]{\includegraphics[scale=0.15]{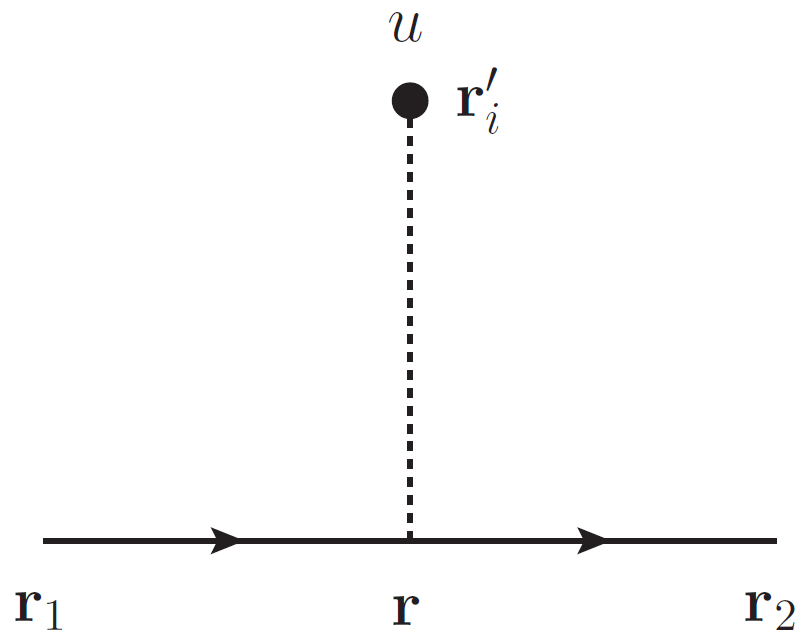}}
	\subfigure[\label{fig:gugug}]{\includegraphics[scale=0.15]{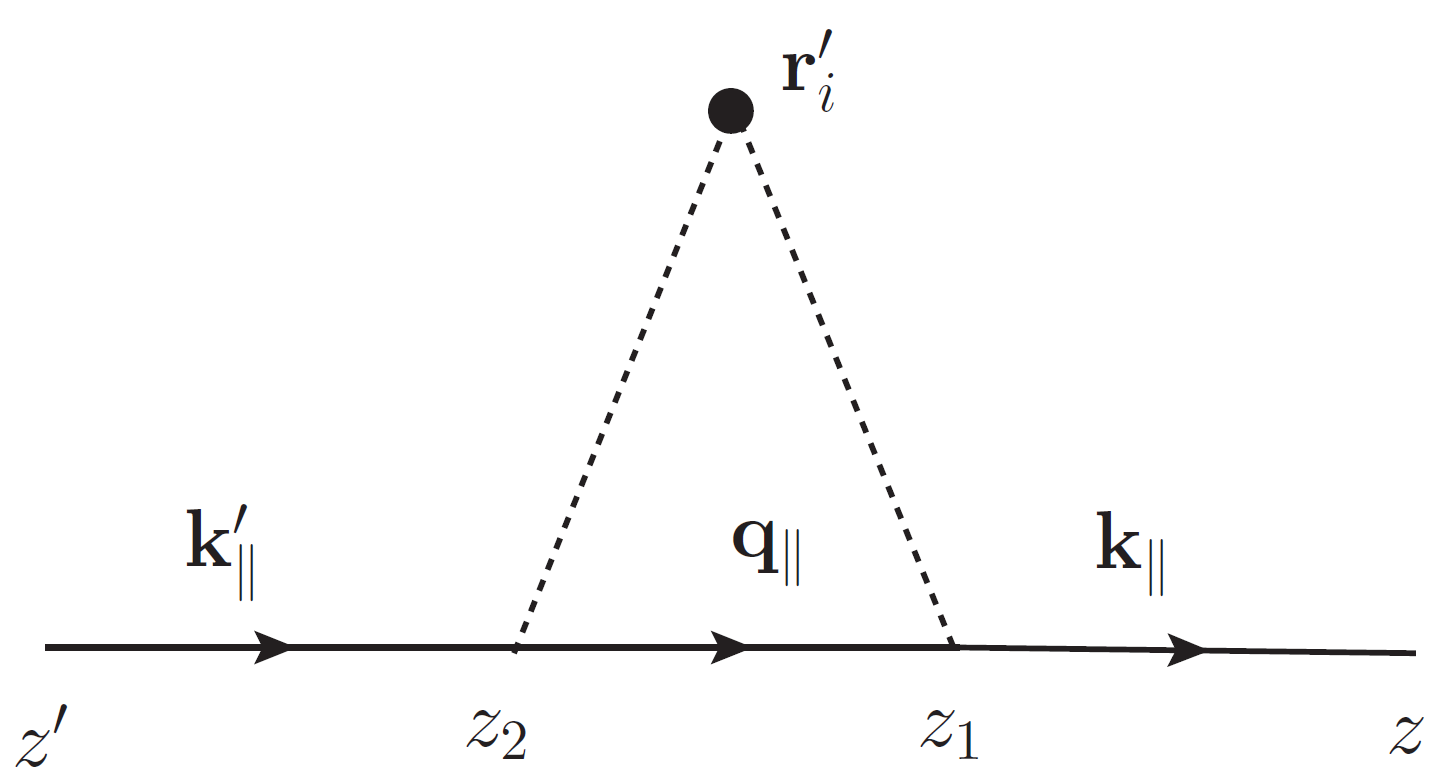}}
	\caption{Feynman diagrams for (a) the first order and (b) the second order corrections to the disordered-averaged Green function. The solid line denotes a free Green function. The dash line and the solid dot represent an impurity potential}
\end{figure}

In this section, we calculate the surface scattering rate, which is related to the imaginary part of the surface self-energy, using the technique from Ref. \cite{Gorbar2016}. The full surface Green function is assumed to have a form
\begin{align} 
G_s(\omega,\vec k_{\|},z,z') = \ell(k_{x}) \frac{\mathbbm{1} - p\sigma_x}{\omega - p v_y k_{y} - \Sigma_s}e^{-\ell(k_{x})(z+z')}, \nonumber \\
\end{align}
where $\Sigma_s$ is the surface self-energy. This Green function can be expanded to the first order in self-energy as
\begin{align} \label{eq:G_s_expand}
G \approx G^0_s + \frac{e^{-\ell(k_{x})(z+z')}}{2\ell(k_{x})} G^0_s\Sigma G^0_s.
\end{align}
Next, we compute the disordered-averaged Green function, $\langle G_s \rangle$, to a certain order in the perturbation theory and then compare it to the right-hand side of Eq. \ref{eq:G_s_expand} to extract the surface self-energy. The first order correction to the surface Green function which describes a particle scatters off an impurity (located at $\vec r'_i$) at lowest order can be calculated, according to Fig. \ref{fig:gug}, as
\begin{align} \label{eq:F_one_impurity}
	G_s^{(1)}(\omega,\vec r_1,\vec r_2,\vec r'_i) = \int d^3r & G_s^0(\omega,\vec r - \vec r_1)u(\vec r - \vec r'_i) \nonumber \\
	&\times G_s^0(\omega,\vec r_2 - \vec r).
\end{align}
Summing over all impurities and performing the disordered average, the first order correction to the surface Green function has a form
\begin{align}
\langle& G^{(1)}_s(\omega,\vec k_{1\|},\vec k_{2\|},z_1,z_2) \rangle_{\mathrm{dis}} = (2\pi)^2\delta(\vec k_{1\|} - \vec k_{2\|})n_\mathrm{imp}u_0 \nonumber \\
&\times\frac{e^{-\ell(k_x)(z_1 + z_2)}}{2\ell(k_{x})}G_s^0(\omega,\vec k_{1\|},z,z_1)G_s^0(\omega,\vec k_{1\|},z_2,z).
\end{align}
Upon comparing to Eq. \ref{eq:G_s_expand}, one finds the first order in the surface self-energy equals a constant,
\begin{align}
\Sigma^{(1)} = n_\mathrm{imp} u_0,
\end{align}
where $n_{\mathrm{imp}} = N/V$ is an impurity concentration. For the Green function in the grand canonical ensemble, this first order in self-energy is simply a shift in a chemical potential: $\mu' = \mu - n_\mathrm{imp} u_0$.

For the second order correction, the one-particle-irreducible Green function (Fig. \ref{fig:gugug}) can be calculated as
\begin{widetext}
\begin{align}
G_s^{(2)}(\omega,\vec k_{\|},\vec k'_{\|},\vec r'_{i,\|},z,z',z'_i) = & \int dz_1 dz_2 \int \frac{d^2q_{\|}}{(2\pi)^2}  G_s^0(\omega,\vec k_{\|},z,z_1)u_0\delta(z'_i-z_1)e^{-i\vec r'_{i,\|}\cdot(\vec k_{\|}-\vec q_{\|})}G^0(\omega,\vec q_{\|},z_1,z_2)u_0\delta(z'_i-z_2) \nonumber \\
& \tabi \ \ \ \ \ \ \ \ \times e^{-i\vec r'_{i,\|}\cdot(\vec q_{\|}-\vec k'_{\|})}G_s^0(\omega,\vec k'_{\|},z_2,z').
\end{align}
\end{widetext}
Performing the disorder average, we have
\begin{align} \label{eq:Gs2}
&\langle G_s^{(2)}(\omega,\vec k_{\|},\vec k'_{\|},z,z') \rangle_{\mathrm{dis}} = (2\pi)^2\delta(\vec k_{\|} - \vec k'_{\|}) n_\mathrm{imp}u_0^2 \nonumber \\
&\times \int\frac{d^2q_{\|}}{(2\pi)^2} \int dz'_i G_s^0(\omega,\vec k_{\|},z,z'_i)G^0(\omega,\vec q_{\|},z'_i,z'_i) \nonumber \\
&\tabi \ \ \ \ \times G_s^0(\omega,\vec k'_{\|},z'_i,z').
\end{align}
Depending on scattering processes one considers, the intermediate fermion line with parallel momentum $\vec q_{\|}$ in Fig. \ref{fig:gugug} can be replaced by either surface or bulk Green functions.

It is often convenient to regard the disordered average of the second-order diagram as an effective electron-electron interaction. This interaction is captured by the correlation function,
\begin{align}
	D(\vec r_1 - \vec r_2) &\equiv	\left\langle \sum_{i} u(\vec r_1 - \vec r'_i) u(\vec r_2 - \vec r'_i)\right\rangle_{\mathrm{dis}} \nonumber \\
	&= n_{\mathrm{imp}} u_0^2 \delta(\vec r_1 - \vec r_2).
\end{align}
The corresponding term in the Hamiltonian is $H_D = \int  \psi^\dagger(\vec r) \psi(\vec r) D(\vec r - \vec r') \psi^\dagger(\vec r') \psi(\vec r') d^3rd^3r'$. We will use this point of view later in the calculations involving bulk states and vertex corrections.

\begin{figure}
	\centering
	\subfigure[\label{fig:gammasts}]{\includegraphics[scale=0.20]{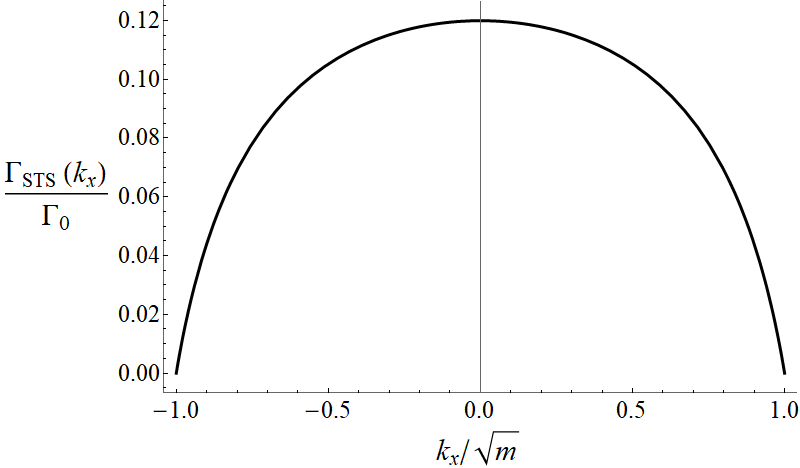}}
	\subfigure[\label{fig:gammab0}]{\includegraphics[scale=0.20]{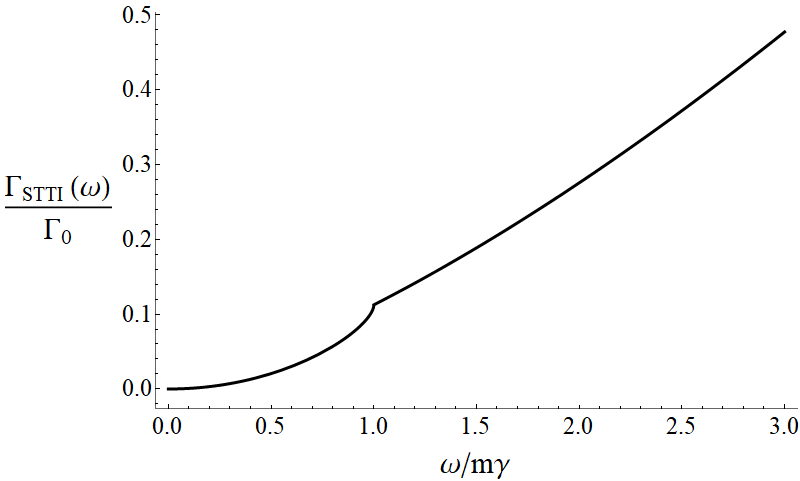}}
	\caption{Plots of surface scattering rates for (a) STS process and (b) Surface-to-translationally-invariant-bulk (STTI) process. The parameter $\Gamma_0 \equiv \frac{n_{imp}u_0^2 m^{3/2}}{v_y v_z}$ has a units of a scattering rate.}
\end{figure}

In the case of the surface-to-surface (STS) scattering, we set $G^0$ in Eq. \ref{eq:Gs2} to be $G_s^0$ and then compare the result with Eq. \ref{eq:G_s_expand}. We find the contribution to the surface self-energy from the STS scattering to be
\begin{align} \label{eq:sigma_s_2s}
\Sigma_{STS}^{(2)}(\omega,k_x) &=& n_\mathrm{imp}u_0^2\int \frac{d^2q_{\|}}{(2\pi)^2}\frac{1}{\omega-pv_yq_{y}} \frac{2\ell(k_{x})\ell(q_{x})}{\ell(k_{x})+\ell(q_{x})}. \nonumber \\
\end{align}
One can perform an integral over $\vec q_{\|}$ analytically (see Ref. \cite{Gorbar2016} or Appendix \ref{app:sigma_s}). The surface scattering rate due to the STS process can then be obtained from the imaginary part of the retarded self-energy as
\begin{align}
\Gamma_{STS}(k_x) =& - Im \Sigma_{STS}^{(2)}(\omega+i\eta,k_x) \nonumber \\
=& \frac{n_\mathrm{imp}u_0^2}{\pi v_y }\ell(k_x) \nonumber \\
\times \Bigg[\sqrt{m}  -& \frac{m-k_{x}^2}{\sqrt{2m-k_{x}^2}}\tanh^{-1}\left(\frac{\sqrt{m}}{\sqrt{2m-k_{x}^2}}\right) \Bigg].
\end{align}
The plot of the scattering rate $\Gamma_{STS}$ as a function of $k_x$ is displayed in Fig. \ref{fig:gammasts}. The rate is peaked at $k_x = 0$ and diminishes as $k_x \rightarrow \pm \sqrt{m}$. 

In the case of the surface-to-bulk (STB) scattering, the intermediate fermion line in Fig. \ref{fig:gugug} is substituted by the bulk Green function $G_b^0$. As shown in Sec. \ref{sec:green_function}, there are two contributions to the bulk Green function in the presence of an interface: the translationally invariant part, $G_{ti}$, and the nontranslationally invariant part, $G_{ni}$. The STB scattering needs to include the processes in which the surface states scatter into both of these bulk contributions, namely, the surface-to-translationally-invariant-bulk (STTI) and surface-to-nontranslationally-invariant-bulk (STNI) processes. Let us first consider the case of the STTI scattering process. Substituting $G^0$ in Eq. \ref{eq:Gs2} by $G^{(0)}_{ti}$, we have
\begin{align}
&\langle G_{STTI}^{(2)}(\omega,\vec k_{\|},\vec k'_{\|},z,z') \rangle_{\mathrm{dis}} = (2\pi)^2\delta(\vec k_{\|} - \vec k'_{\|}) n_\mathrm{imp}u_0^2 \nonumber \\
&\times \int\frac{d^3q}{(2\pi)^3} \int dz'_i G_s^0(\omega,k_{\|},z,z'_i)G^0_{ti}(\omega,\vec q)G_s^0(\omega,k'_{\|},z'_i,z'),
\end{align}
where $G_{ti}^0(\omega,\vec q)$ is given in Eq. \ref{eq:Gb0_2}. From the identity
\begin{align}  \label{eq:sigma}
(1-p\sigma_x)\sigma_i(1-p\sigma_x) = \begin{cases}
	0 &\mathrm{if} \ i \neq x\\
	-p(1-p\sigma_x)^2 &\mathrm{if} \ i = x,
\end{cases}
\end{align}
one finds that only the coefficients of $\mathbbm{1}$ and $\sigma_x$ in the integrand are nonzero. Integrating over $z'_i$ and then comparing $\langle G_{STTI}^{(2)} \rangle_{\mathrm{dis}}$ with Eq. \ref{eq:G_s_expand}, one concludes that the retarded self-energy is
\begin{align} \label{eq:sigma_s_b0}
\Sigma_{STTI}^{(2)}(\omega) = n_\mathrm{imp}u_0^2 \omega \int\frac{d^3q}{(2\pi)^3} \frac{1}{\omega^2 - \varepsilon_q^2 + i\eta\mathrm{sgn}(\omega)}.
\end{align}
Using the identity,
\begin{align} \label{eq:principal_delta}
	\frac{1}{\omega^2 - \varepsilon_q^2+i\eta\mathrm{sgn}(\omega)} = P\frac{1}{\omega^2 - \varepsilon_q^2} - i\pi\mathrm{sgn}(\omega)\delta(\omega^2 - \varepsilon_q^2), \nonumber \\
\end{align}
we obtain the imaginary part of the retarded self-energy as
\begin{align} \label{eq:sigma_b0_a}
\mathrm{Im} \Sigma_{STTI}^{(2)} = -\pi n_\mathrm{imp}u_0^2 |\omega| \int\frac{d^3q}{(2\pi)^3}\delta(\omega^2 - \varepsilon_q^2).
\end{align}
The integral in Eq. \ref{eq:sigma_b0_a} can be performed analytically (see Ref. \cite{Gorbar2016} or Appendix \ref{app:sigma_s}) and, thus, the scattering rate is given by
\begin{align} \label{eq:sigma_b0_final}
&\Gamma_{STTI}(\omega) = \frac{n_\mathrm{imp}u_0^2 |\omega|}{4\pi v_y v_z} \nonumber \\
&\times \left( \sqrt{m + \frac{|\omega|}{\gamma}} -\theta\left(m - \frac{|\omega|}{\gamma}\right)\sqrt{m - \frac{|\omega|}{\gamma}} \right).
\end{align}
The plot of $\Gamma_{STTI}$ vs. $\omega$ is displayed in Fig. \ref{fig:gammab0}. We find that, as frequency $\omega$ increases, the scattering rate of the $STTI$ process increases. We can understand this behavior from the fact that there is more scattering phase space at higher energy. The kink located at $\omega = m\gamma$ can be attributed to the Lifshitz transition, at which point the rate of change of the density of states with respect to chemical potential is discontinuous.

\begin{figure}
	\centering
	\subfigure[\label{fig:gammacompare1}]{\includegraphics[scale=0.25]{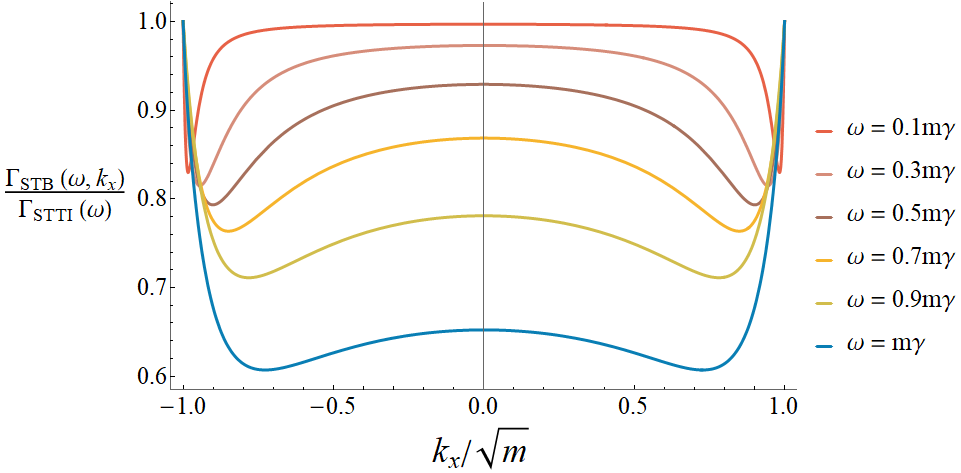}}
	\subfigure[\label{fig:gammacompare2}]{\includegraphics[scale=0.25]{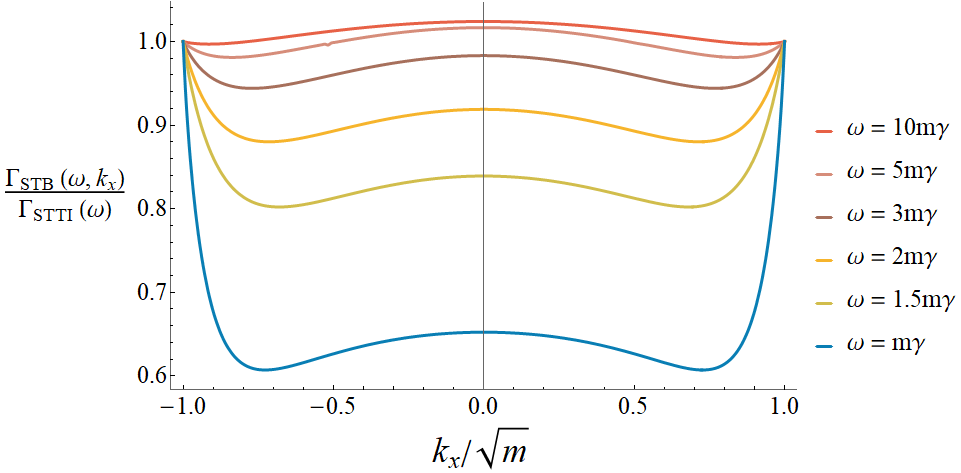}}
	\subfigure[\label{fig:gammabk0}]{\includegraphics[scale=0.25]{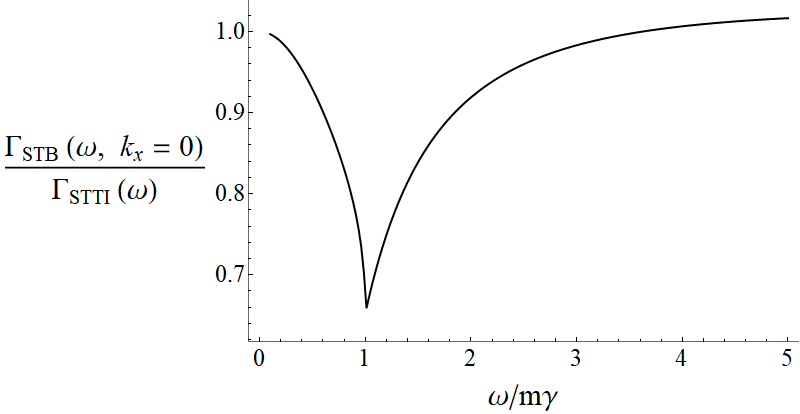}}
	\caption{Plots of the ratios between the scattering rate of the total STB process to that of the STTI process (a,b) as a function of $k_x$ at various $\omega$ and (c) as a function of $\omega$ at $k_x = 0$.} \label{fig:gammastb} 
\end{figure}

We next calculate the contribution to the self-energy from the STNI process. Following the same procedure as the case of $G_{ti}$, we find the self-energy to be
\begin{align}
\Sigma_{STNI}^{(2)} (\omega,k_x)
= & n_\mathrm{imp}u_0^2 \omega \int\frac{d^3q}{(2\pi)^3} \frac{1}{\omega^2 - \varepsilon_q^2 + i\eta\mathrm{sgn}(\omega)} \nonumber \\
& \times\frac{(\tilde{q}_z + i\ell(q_{x}))}{(\tilde{q}_z- i\ell(q_{x}))} \frac{\ell(k_{x})}{(\ell(k_{x}) - i\tilde{q}_z)}.
\end{align}
Using Eq. \ref{eq:principal_delta}, one obtains the imaginary part of the retarded self-energy\footnote{The Cauchy principal value is real and the integral over the delta function term is pure imaginary. One can easily check these assertions by taking a complex conjugate of the integral and then reverse the sign of $\tilde{q}_z$.} as
\begin{align} \label{eq:sigma_b1_a}
\mathrm{Im}\Sigma_{STNI}^{(2)}(\omega,k_x) = &-\pi n_\mathrm{imp}u_0^2 |\omega| \int\frac{d^3q}{(2\pi)^3} \frac{(\tilde{q}_z + i\ell(q_{x}))}{(\tilde{q}_z- i\ell(q_{x}))} \nonumber \\ 
&\times \frac{\ell(k_{x})}{(\ell(k_{x}) - i\tilde{q}_z)} \delta(\omega^2 - \varepsilon_q^2).
\end{align}
Combining the scattering rates form the STTI and STNI proccesses, $\Gamma_{STB} = -\mathrm{Im}\Sigma_{STTI}^{(2)}(\omega+i\eta)-\mathrm{Im}\Sigma_{STNI}^{(2)}(\omega+i\eta,k_x)$, and intergrating over $q_y$, we arrive at the expression for the total STB scattering rate,
\begin{align} \label{eq:sigma_b_total}
\Gamma_{STB}&(\omega,k_x) = \frac{n_\mathrm{imp}u_0^2 |\omega|}{8\pi^2 v_y} \int \frac{\theta(\omega^2 - v_z^2(\tilde{q}_z^2 + \ell^2(q_{x})))}{\sqrt{\omega^2 - v_z^2(\tilde{q}_z^2 + \ell^2(q_{x}))}} \nonumber \\ 
&\times \frac{\tilde{q}_z^2(\tilde{q}_z^2 + (\ell(q_{x})-\ell(k_{x}))^2 + \ell^2(k_{x}))}{(\tilde{q}_z^2+\ell^2(q_{x}))(\tilde{q}^2_z + \ell^2(k_{x}))} dq_{x}d\tilde{q}_z.
\end{align}
Note that $\Gamma_{STB}$ vanishes at $\omega = 0$. This is an indication that the surface states are decoupled from the bulk states at this value of energy. To understand the behavior of $\Gamma_{STB}$, we plot the ratio $\Gamma_{STB}/\Gamma_{STTI}$ in Fig. \ref{fig:gammastb}. Since the total STB scattering is the sum of the STTI and STNI processes, the ratio $\Gamma_{STB}/\Gamma_{STTI} = 1$ if STNI does not affect the scattering rate. The deviation of $\Gamma_{STB}/\Gamma_{STTI}$ from unity indicates how much the STNI process contributes to the total $\Gamma_{STB}$. From Fig. \ref{fig:gammacompare1}, one can see that, for a wide range of $k_x$, the ratio decreases as $\omega$ increases from $0$ to $m\gamma$. However, the ratio starts to bounce up, as $\omega$ increases beyond $m\gamma$, and, eventually, reaches $1$ at large $\omega$ (Fig. \ref{fig:gammacompare2}). This behavior can be clearly illustrated in a plot of $\Gamma_{STB}/\Gamma_{STTI}$ vs. $k_x$ at $\omega = 0$ (Fig. \ref{fig:gammabk0}). Consequently, including the STNI process results in a lowering of the total bulk scattering rate. 

\begin{figure}
	\centering
	\includegraphics[scale=0.25]{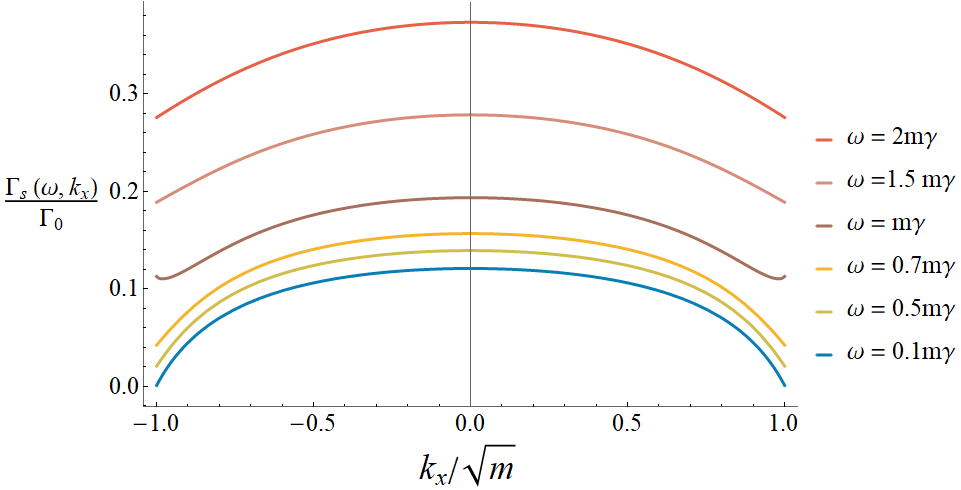}
	\caption{Plot of total surface scattering rate $\Gamma_{s}$ as a function of $k_x$ at various frequencies.} \label{fig:gammas}
\end{figure}

Finally, the total surface scattering rate can be calculated from,
\begin{align} \label{eq:gammas}
	\Gamma_{s}(\omega,k_x) = \Gamma_{STS}(k_x) + \Gamma_{STB}(\omega,k_x).
\end{align}
The plot of $\Gamma_s$ vs. $k_x$ is displayed in Fig. \ref{fig:gammas}. $\Gamma_s$ has a maximum value at $k_x = 0$ and, unlike the case of $\Gamma_{STS}$, $\Gamma_s$ decreases to some finite values at $k_x = \pm\sqrt{m}$. 

\section{Surface magnetoelectric susceptibility} \label{sec:surface_response}

In this section, we calculate the Edelstein effect response or the surface states' magnetoelectric susceptibility, $\chi_{ij}$, using linear response theory. In general, $\chi_{ij}$ is defined through 
\begin{align}
M_{i}(\vec r, t) = \int \chi_{ij}(\vec r, \vec r'; t, t') E_{j}(\vec r',t') d^3r' dt',
\end{align}
where $\vec M$ is the induced magnetization and $\vec E$ is the applied electric field. In order to compute $\chi$, one considers an action of the form
\begin{align} \label{eq:action_full}
S[\vec A,\vec B] = S_0[\vec A] - \int \vec M \cdot \vec B
\end{align}
where $S_0[\vec A]$ is an unperturbed action, $\vec A$ is a vector potential, $\vec B$ is an external magnetic field, and $\vec M = \mu_B\psi^\dagger \vec \sigma\psi$ is a magnetization with $\mu_B$ being the Bohr magneton. In this paper, we will calculate $\chi$ using the grand canonical ensemble. This means the action $S_0$ needs to include a chemical potential $\mu$. 

The outline of the calculation for the magnetoelectric susceptibility is as follows. Let $\mathcal{Z[\vec A,\vec B]} = \int D[\psi, \psi^\dagger] e^{-S[\vec A,\vec B]}$ be the partition function of the action in Eq. \ref{eq:action_full}. First, we compute the response function $L_{ij}$ within the imaginary time formalism \cite{Altland2010} from
\begin{align} 
L_{ij}(\vec r,\vec r',\tau,\tau') =& -\frac{1}{Z}\frac{\delta^2}{\delta B_i(\vec r,\tau) \delta A_j(\vec r',\tau')}\Bigg|_{A,B=0}\mathcal{Z}[\vec A,\vec B] \nonumber \\
=& \frac{1}{Z}\int D[\psi,\psi^\dagger] M_i J_j e^{-S}
\end{align}
where $J_i \equiv \frac{\delta S_0}{\delta A_i}$ is a $U(1)$ current. Next, we perform the disordered average on the response function. For simplicity of notation, we define the bracket symbol to include both the thermal and disordered averages. Hence, the resulting $L_{ij}$ is given by
\begin{align} \label{eq:Lij}
L_{ij} = & \langle M_i J_j \rangle.
\end{align}
The response $L_{ij}$ is then calculated as a function of Matsubara frequency, $\omega_n$, and momentum $\vec q$. In the case of surface states, $\vec q_{\|}$ and the coordinates perpendicular to the surface, $z'$, replace $\vec q$ as independent variables in $L_{ij}$. Once $L_{ij}$ is analytic continued to real frequencies $i\omega_n \rightarrow \omega + i\eta$, the Edelstein effect response can be obtained from $L_{ij}$ as
\begin{align} \label{eq:Ltochi}
\chi_{ij}(\omega, \vec q) = \frac{i}{\omega}L_{ij}(\omega,\vec q).
\end{align}

The unperturbed action of the Weyl semimetal model we consider in this paper (Eq. \ref{eq:weyl_hamiltonian}) is $S_0[\vec A] = \int d^3r d\tau \mathcal{L}_0[\vec A]$ with the Lagrangian given by
\begin{align}
&\mathcal{L}_0 = \psi^\dagger (\partial_\tau  - \mu)\psi + \psi^\dagger p\gamma[(-i\partial_x - eA_x)^2 - m]\sigma_y \psi  \nonumber \\
&+ \psi^\dagger v_z\sigma_z(-i\partial_z - eA_z)\psi - \psi^\dagger v_y(-i\partial_y - eA_y)\sigma_x\psi \nonumber \\
&+ \psi^\dagger U \psi
\end{align}
Here, as mentioned above, there is a chemical potential $\mu$ in $\mathcal{L}_0$, because we plan to calculate $\chi$ within the grand canonical ensemble. The inclusion of $\mu$ results in a shift in the frequency dependence of the Green function from $G(\omega)$ to $G(\omega + \mu)$. From the action $S_0[\vec A]$, we calculate the $U(1)$ current as
\begin{align}
J_x(\vec r,t) = \frac{\delta S_0[\vec A]}{\delta A_x(\vec r,t)} =& -2e p \gamma \psi^\dagger \sigma_y (-i\partial_x - eA_x)\psi(\vec r,t), \label{eq:Jx}  \\
J_y(\vec r,t) = \frac{\delta S_0[\vec A]}{\delta A_y(\vec r,t)} =& ev_y \psi^\dagger \sigma_x \psi(\vec r,t), \label{eq:Jy} \\
J_z(\vec r,t) = \frac{\delta S_0[\vec A]}{\delta A_z(\vec r,t)} =& -e v_z \psi^\dagger \sigma_z \psi(\vec r,t). \label{eq:Jz} 
\end{align}
Eq. \ref{eq:Lij} and the forms of $G_s$, $\vec M$, and $\vec J$ imply the matrix structure of $L_{ij}$ is
\begin{align} \label{eq:Lij_matrix}
L_{ij} \sim \mathrm{tr}( \sigma_i (1-p\sigma_x) \sigma_k (1-p\sigma_x) ), 
\end{align}
where the index $k = \begin{cases}
	x &\mathrm{if} \  j = y \\
	y &\mathrm{if} \  j = x \\
	z &\mathrm{if} \  j = z
\end{cases}$. Using Eq. \ref{eq:sigma} and the cyclical property of trace, we find that only the component $xy$ is nonvanishing. Physically, $L_{xy}$ is a magnetization response along the $x$ direction due to an applied current along the $y$ direction.

\begin{figure}
	\centering
	\subfigure[\label{fig:vertexfzz}]{\includegraphics[scale=0.25]{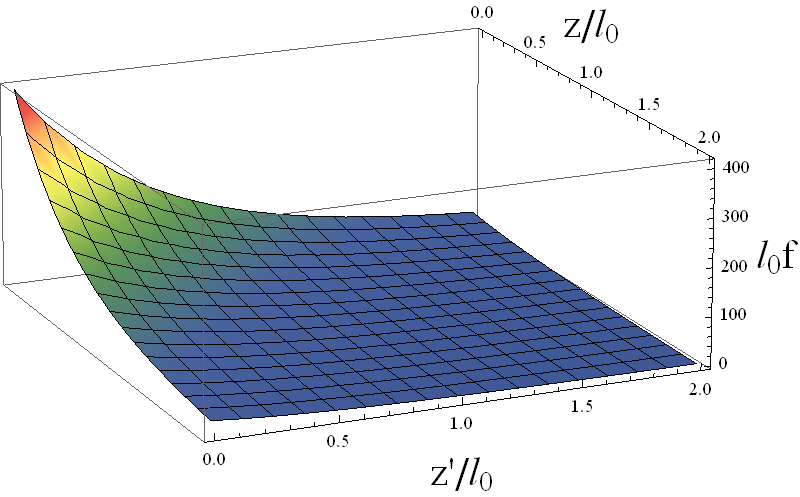}}
	\subfigure[\label{fig:vertexerror}]{\includegraphics[scale=0.25]{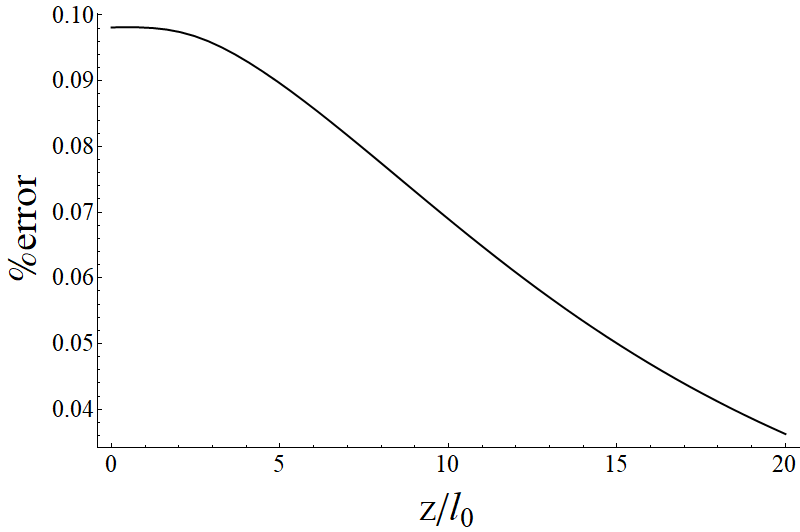}}
	\caption{Vertex correction, $f(\omega,z,z')$, obtained from solving Eq. \ref{eq:fredholm} numerically with $\omega$ set to $0.1 m\gamma$. Shown in (a) is the three-dimensional plot of $f(\omega,z,z')$. In (b), the percentage difference between $f(\omega,z,z')$ and right-hand sides of Eq. \ref{eq:fredholm} is plotted against $z = z'$. Here, the parameter $l_0 \equiv v_z/m\gamma$ has a unit of length.} \label{fig:vertexresult}
\end{figure}

Substituting $M_x$ and $J_y$ into Eq. \ref{eq:Lij}, we have $L_{xy} = ev_y \mu_B \langle \psi^\dagger \sigma_x \psi \psi^\dagger \sigma_x \psi \rangle$. Using Wick's theorem, we find
\begin{align}
&L_{xy}(i\omega_n,\vec r) = -ev_y\mu_B T\sum_{l} \int d^3r' \nonumber \\ 
&\times \mathrm{tr}( \langle \sigma_x G(i\omega_l + \mu,\vec r,\vec r')\sigma_x G(i\omega_l - i\omega_n + \mu,\vec r',\vec r) \rangle_{\mathrm{dis}}).
\end{align}
Here, the negative sign in the front comes from the fermion loop. The effect of the disordered average can be captured by the surface states' vertex $\Lambda_{s}$ as
\begin{align} \label{eq:Lxy}
&L_{xy}(i\omega_n,\vec r) = -ev_y\mu_B T\sum_{l} \int d^3r' d^3r_1d^3r_2 \nonumber \\
&\times \mathrm{tr}\left[\sigma_x G(i\omega_l + \mu,\vec r,\vec r_1) \Lambda_{s}(i\omega_l + \mu,i\omega_l - i\omega_n + \mu,\vec r_1, \vec r', \vec r_2) \right. \nonumber \\
& \ \ \ \ \ \ \left. G(i\omega_l - i\omega_n + \mu,\vec r_2,\vec r)\right].
\end{align}
We follow the standard procedure by converting the Matsubara summation to a contour integral and then performing an analytic continuation on $L_{xy}(i\omega_n)$ to real frequencies. We find the retarded response function is given by
\begin{widetext}
\begin{align}\label{eq:Lxy_s}
L_{xy}(\omega',\vec q_{\|},z) = & \frac{ev_y\mu_B}{2\pi i} \int_{-\infty}^{\infty}d\omega\left(n_F(\omega-\omega'-\mu) - n_F(\omega-\mu)\right) \int\frac{d^2k_{\|}}{(2\pi)^2}\int dz'dz_1dz_2 \nonumber \\
&\times\mathrm{tr}\left(\sigma_x G^A_s(\omega,\vec k_{\|},z,z_1) \Lambda_{s}(\omega,\omega - \omega',\vec k_{\|},\vec k_{\|} - \vec q_{\|}, z_1, z', z_2) G^R_s(\omega - \omega',\vec k_{\|}-\vec q_{\|}, z_2,z)\right),
\end{align}
\end{widetext}
where $G^A$ and $G^R$ are the advanced and retarded Green functions, respectively.
\begin{figure}
	\includegraphics[scale=0.35]{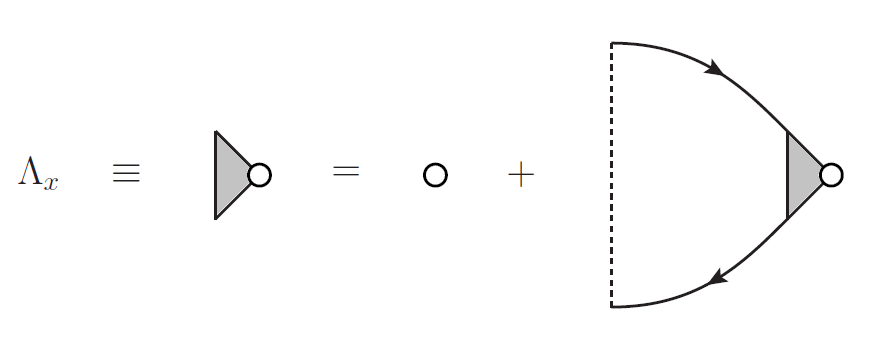}
	\caption{Vertex function in the self-consistent ladder approximation. The white dot represents $\sigma_x$ and the shaded triangle denotes the vertex $\Lambda_x$.} \label{fig:vertex}
\end{figure}
From Eq. \ref{eq:Lxy_s}, we take the limit $\vec q_{\|}\rightarrow 0$, $\omega'\rightarrow 0$, and use Eq. \ref{eq:Ltochi}. We find the surface magnetoelectric susceptibility is
\begin{align} \label{eq:chi_s}
	\chi_{s}(z,\mu,T) & \equiv   \chi_{xy}(z,\mu,T)\nonumber \\
	& = \int_{-\infty}^{\infty}d\omega \left(-\frac{\partial n_F(\omega - \mu)}{\partial \omega}\right) X_s(\omega,z) 
\end{align}
where the function $X_s(\omega,z)$ is 
\begin{align} \label{eq:X_s}
&X_s(\omega,z) = \frac{ev_y\mu_B}{2\pi}\int\frac{d^2k_{\|}}{(2\pi)^2}\int dz'dz_1dz_2  \nonumber \\
&\times\mathrm{tr}\left[\sigma_x G^A_s(\omega,\vec k_{\|},z,z_1) \right. \nonumber \\
&\left. \ \ \ \ \ \ \ \Lambda_{s}(\omega,\vec k_{\|}, z_1, z', z_2) G^R_s(\omega,\vec k_{\|}, z_2,z)\right].
\end{align}
Here, the vertex $\Lambda_{s}(\omega,\vec k_{\|},z_1,z',z_2)$ is a shorthand for $\Lambda_{s}(\omega,\omega,\vec k_{\|},\vec k_{\|},z_1,z',z_2)$. The surface Green function is given by 
\begin{align} \label{eq:Gs_full_AR}
G^{R,A}_s(\omega,\vec k_{\|}, z, z') = \frac{\ell(k_{x})(\mathbbm{1} - p\sigma_x)e^{-(z+z')\ell(k_{x})}}{\omega - pv_y k_{y} \pm i\Gamma_s(\omega,k_{x})},
\end{align}
where, in the denominator, the positive sign is for the retarded Green function, $G^R$, and the negative sign is for the advanced Green function, $G^A$.  Within the self-consistent ladder approximation (Fig. \ref{fig:vertex}), the vertex function, $\Lambda_{s}(\omega,\vec k_{\|},z_1,z',z_2)$, can be obtained by solving the equation,
\begin{align} \label{eq:vertex_ladder}
&\Lambda_{s}(\omega,\vec k_{\|},z_1,z',z_2) = \sigma_x \delta(z_1 - z')\delta(z' - z_2) \nonumber \\
& +  \int \frac{d^2 s_{\|}}{(2\pi)^2}\int dz_1'\int dz_2' D(\vec k_{\|} - \vec s_{\|}, z_1, z_2)  \nonumber \\
&\times G^A_s(\omega,\vec s_{\|}, z_1,z_1') \Lambda_{s}(\omega,\vec s_{\|},z_1',z',z_2') G^R_s(\omega,\vec s_{\|},z_2',z_2).
\end{align}
We make an ansatz,
\begin{align} \label{eq:Lambda_ansatz}
	&\Lambda_{s}(\omega,\vec k_{\|},z_1,z',z_2) =  \sigma_x\delta(z_1 - z')\delta(z'-z_2) \nonumber \\
	& + \delta(z_1 - z_2) \frac{\mathbbm{1} -p\sigma_x}{2}(-p)f(\omega,z_1,z'),
\end{align}
that the vertex is the sum of a bare vertex and a vertex correction. Plugging in this ansatz, the Green functions (from Eq. \ref{eq:Gs_full_AR}), and $D(\vec k_{\|} - \vec s_{\|},z_1,z_2) = n_{\mathrm{imp}}u_0^2\delta(z_1 - z_2)$ into Eq. \ref{eq:vertex_ladder}, and then
comparing both sides of the equation, we find that $f(\omega,z_1,z_2)$ satisfies the integral equation,
\begin{align} \label{eq:fredholm}
f(\omega,z_1,z') = F(\omega,z_1,z') + \int dz_1'F(\omega,z_1,z_1')f(\omega,z_1',z'),
\end{align}
where the function $F(\omega,z,z')$ is defined by
\begin{align} \label{eq:suface_F}
F(\omega,z,z') \equiv \frac{2n_{\mathrm{imp}}u_0^2}{v_y}\int_{-\sqrt{m}}^{\sqrt{m}} \frac{ds_{x}}{2\pi} \frac{\ell^2(s_x)e^{-2(z+z')\ell(s_{x})}}{\Gamma_s(\omega,s_{x})}.
\end{align}
Eq. \ref{eq:fredholm} is a Fredholm integral equation of the second kind. We implement the standard quadrature algorithm from Ref. \cite{Rahbar2008} to numerically solve Eq. \ref{eq:fredholm}. As an example, the solution in the case of $\omega = 0.1 m\gamma$ is displayed in Fig. \ref{fig:vertexfzz}. In computing the quadrature, we choose an uneven grid spacing. That is the grids are chosen to be dense at small $(z,z')$ and then the spacing becomes wider as $(z,z')$ increase. The grid sizes we use are small enough such that the difference between the left- and right-hand sides of Eq. \ref{eq:fredholm} is less than $0.1\%$ at $\omega = 0.1m\gamma$ (see Fig.\ref{fig:vertexerror}). 

\begin{figure}
	\centering
	\includegraphics[scale=0.30]{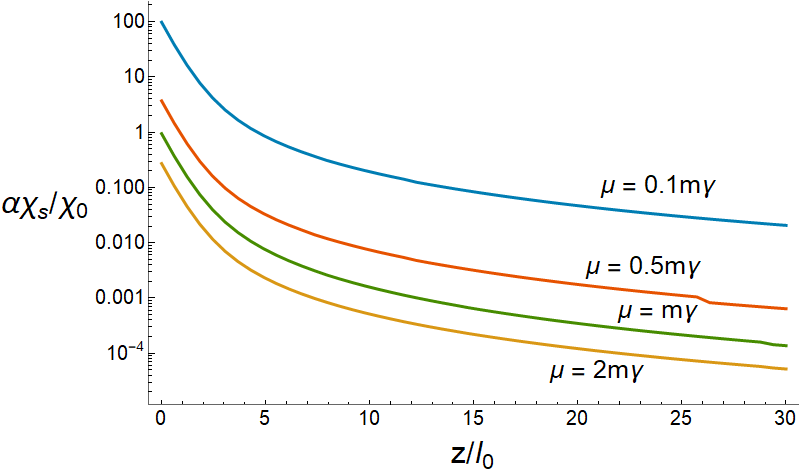}
	\caption{Plots of surface states' magnetoelectric susceptibilities ($\chi_s$) as a function of a distance from the surface ($z$) at four values of chemical potentials. The temperature is set to be $T = 0$.} \label{fig:chis}
\end{figure}

Substituting $\Lambda_x$ from Eq. \ref{eq:Lambda_ansatz} into Eq. \ref{eq:X_s} and simplifying the expression, one finds
\begin{align} \label{eq:XS_total}
X_s(\omega,z) = X^{(0)}_s(\omega,z) + X^{(1)}_s(\omega,z),
\end{align}
where
\begin{align} \label{eq:XS_0}
X^{(0)}_s(\omega,z)&=\frac{e\mu_B}{4\pi^2}\int^{\sqrt{m}}_{-\sqrt{m}} dk_{x} \frac{\ell(k_{x})e^{-2\ell(k_{x})z}}{\Gamma_s(\omega,k_{x})}
\end{align}
comes from the bare vertex $\sigma_x$ and
\begin{align} \label{eq:XS_1}
X^{(1)}_s(\omega,z)&= \frac{e\mu_B}{2\pi^2}\int^{\sqrt{m}}_{-\sqrt{m}} dk_{x}\int_0^\infty dz' \int_0^\infty dz_1  \nonumber \\ 
	&\times \frac{\ell^2(k_{x})e^{-2(z+z_1)\ell(k_{p,x})}}{\Gamma_s(\omega,k_{x})} f(\omega,z_1,z')
\end{align}
arises from the vertex correction. Since the results above do not depend on the pseudospin index $p$, one needs to multiply $\chi_{s}$ in Eq. \ref{eq:chi_s} by a factor of $2$ to obtain the total susceptibility from the two pseudospin sectors. By changing units of various variables to be dimensionless (see Sec. \ref{app:dimless}), one can express the susceptibility in a scaling form as 
\begin{align} \label{eq:chis_scaling}
\chi_s = 2\frac{\chi_0}{\alpha}\tilde{\chi}_s\left(\frac{z}{l_0},\frac{\mu}{m\gamma},\frac{T}{m\gamma}\right),
\end{align}
where $\tilde{\chi}_s$ is the dimensionless surface susceptibility, the constant $\chi_0 \equiv \frac{e\mu_B \sqrt{m}}{v_z}$ has a unit of magnetization per electric field, $\alpha \equiv \frac{n_{\mathrm{imp}}u_0^2 \sqrt{m}}{v_y v_z}$ is the dimensionless parameter that quantifing the strength of the impurity scattering, and the constant $l_0 \equiv v_z/m\gamma$ has a unit of length. We can clearly see from Eq. \ref{eq:chis_scaling} that the effect of impurity comes out as an overall multiplication factor. To study the behavior of the surface response, we make a plot of $\chi_s$ vs. $z$ at $T = 0$ in Fig. \ref{fig:chis}. We find that $\chi_s$ is largest at the surface ($z = 0$) and then sharply decreases as one goes deeper inside the bulk. This behavior reflects the fact that the surface states are localized near the surface. Additionally, we find $\chi_s$ is strongest when $\mu$ is close to zero (i.e., $\mu$ is located close to the energy level of the Weyl nodes) and becomes smaller as $\mu$ increases. At $\mu = m\gamma$, $\chi_s$ is smaller than its value at $\mu = 0$ by about two orders of magnitude. 

\section{Bulk Magnetoelectric Susceptibility} \label{sec:bulk_response}

To understand how strong the Edelstein response of the surface states is, we need to compare it with that of the bulk states. As shown in Eq. \ref{eq:Gb_Gb0_Gb1}, there are two contributions to the bulk Green function, the translationally invariant part, $G_{ti}$, and nontranslationally invariant part, $G_{ni}$. In this paper, we consider only the former and neglect the latter in the response calculation of the bulk states. That is the free bulk Green function $G_b^0 = G^0_{ti}$ as given in Eq. \ref{eq:Gb0_2}. Furthermore, only the bulk-to-bulk scattering is included. The results obtained from these approximations correspond to the Edelstein effect from the bulk states in the absence of surface.

We first study how the random impurities affect the Green function by calculating the self-energy. The disordered-averaged self-energy at one loop level is 
\begin{align}
\Sigma_b &= \int D(\vec k - \vec q)G_b^0(\omega,\vec q)\frac{d^3q}{(2\pi)^3} \nonumber \\
&= n_{\mathrm{imp}}u_0^2\left(  \int \frac{\omega}{\omega^2 - \varepsilon^2_q} \frac{d^3q}{(2\pi)^3}\mathbbm{1} -   \int \frac{p v_z \ell(q_x)}{\omega^2 - \varepsilon_q^2} \frac{d^3q}{(2\pi)^3} \sigma_y \right) \nonumber
\end{align} 
Using Eq. \ref{eq:principal_delta}, we find the imaginary part\footnote{Here, the imaginary part means the anti-Hermitian part of the matrix.} of the retarded self-energy to be
\begin{align}
\mathrm{Im}\Sigma_b(\omega) = -i \Gamma_{b,0}(\omega) \mathbbm{1} - i p \Gamma_{b,2}(\omega)\sigma_y,
\end{align}
where
\begin{align}
\Gamma_{b,0}(\omega) &= \frac{n_\mathrm{imp}u_0^2 |\omega|}{8\pi^2} \int\delta(\omega^2 - \varepsilon^2_q)  d^3q, \\
\Gamma_{b,2}(\omega) &= -\frac{n_\mathrm{imp}u_0^2 v_z \mathrm{sgn}(\omega)}{8\pi^2}\int \delta(\omega^2 - \varepsilon_q^2)\ell(q_x) d^3q. \label{eq:Gamma_b2}
\end{align}
$\Gamma_{b,0}$ has the same form as the scattering rate from the STTI process (see Eq. \ref{eq:sigma_b0_a}). Hence, from Eq. \ref{eq:sigma_b0_final}, we have
\begin{align}
\Gamma_{b,0}(\omega) = \frac{n_\mathrm{imp}u_0^2 |\omega|}{4\pi v_yv_z}& \left( \sqrt{m + \frac{|\omega|}{\gamma}} \right. \nonumber \\
&\left. - \theta\left(m - \frac{|\omega|}{\gamma}\right)\sqrt{m - \frac{|\omega|}{\gamma}} \right). 
\end{align}
The integral in $\Gamma_{b,2}$ can be analytically calculated using similar technique as the integral in $\Gamma_{STTI}$ (see Appendix \ref{app:tau1}). The result is 
\begin{align}
\Gamma_{b,2}(\omega) &= -\frac{n_\mathrm{imp}u_0^2\gamma\mathrm{sgn}(\omega)}{4\pi v_y v_z}\left[ m\sqrt{m+\frac{|\omega|}{\gamma}} - \frac{\left(m+\frac{|\omega|}{\gamma}\right)^{\frac{3}{2}}}{3} \right. \nonumber \\
& \left. - \theta\left(m - \frac{|\omega|}{\gamma}\right)\left( m\sqrt{m - \frac{|\omega|}{\gamma}} - \frac{\left(m - \frac{|\omega|}{\gamma}\right)^{\frac{3}{2}}}{3} \right) \right] \nonumber \\
\end{align}
The disordered Green function can be obtained from the Dyson equation, $G_b^{-1} = (G_b^{0})^{-1} - \Sigma_b$. Furthermore, we use the approximation that the real part of the self-energy is neglected and, as in Ref. \cite{Rodionov2015}, we drop any imaginary terms in the numerator of the Green function. The result is
\begin{widetext}
\begin{align}
G^{R,A}_b (\omega,\vec k) = \frac{\omega\mathbbm{1} + v_zk_z \sigma_z - v_yk_y\sigma_x - pv_z\ell(k_x)\sigma_y}{\left(\omega\pm i\Gamma_{b,0}(\omega)\right)^2 - \left((v_zk_z)^2 + (v_yk_y)^2 + (v_z\ell(k_x) \pm i\Gamma_{b,2}(\omega))^2\right)},
\end{align}
\end{widetext}
where the positive and negative signs in the denominator are for the retarded Green function, $G^R$, and the advanced Green function, $G^A$, respectively. 

Following a similar matrix structure analysis as in Eq. \ref{eq:Lij_matrix}, we find, in the case of the bulk response, that only the components $yx$, $xy$, and $zz$ of $L_{ij}$ are nonzero. Since we want to compare the surface and bulk results and $xy$ is the only nonvanishing component of the surface response, we focus on the calculation of $L_{xy}$ for the bulk states. We apply the same procedures used to obtain the expression for the surface susceptibility in Eq. \ref{eq:chi_s} to the case of the bulk Green functions and the bulk vertex function. One finds the bulk magnetoelectric susceptibility can be computed from
\begin{align} \label{eq:chib_formula}
\chi_{b} &\equiv  \chi_{xy} = \frac{ev_y\mu_B}{2\pi} \int_{-\infty}^{\infty}d\omega \left(-\frac{\partial n_F(\omega - \mu)}{\partial \omega}\right) \nonumber \\
&\times \int \frac{d^3p}{(2\pi)^3}\mathrm{tr}\left(\sigma_x G^R_b(\omega,\vec p) \Lambda_{b,x}(\omega,\vec p) G^A_b(\omega,\vec p) \right),
\end{align}
and the bulk vertex function, $\Lambda_{b,i}(\omega,\vec p)$, is obatined by solving the self-consistent equation (see Fig. \ref{fig:vertex}),
\begin{align} \label{eq:vertex_bulk_eq}
\Lambda_{b,i}(\omega,\vec k) = \sigma_i + &\int \frac{d^3q}{(2\pi)^3} D(\vec k - \vec q)  \nonumber \\
&\times G^R_b(\omega,\vec q)\Lambda_{b,i}(\omega,\vec q)G^A_b(\omega,\vec q).
\end{align}
To solve this equation, we make an ansatz in a similar fashion as the surface case (Eq. \ref{eq:Lambda_ansatz}), 
\begin{align} \label{eq:ansatz_bulk}
\Lambda_{b,i} = \sigma_i + f^0_i\mathbbm{1} + \vec f_i \cdot \vec \sigma.
\end{align}
Substituting Eq. \ref{eq:ansatz_bulk} into Eq. \ref{eq:vertex_bulk_eq} leads to the system of self-consistent equations for the functions $f^0_i$ and $\vec f_i = (f_i^x,f_i^y,f_i^z)$. In the case of $\Lambda_{b,x}$, the solutions to the self-consistent equation for $\vec f_x$ (see Appendix \ref{app:bulk_vertex_selfcon}) are
\begin{align}
&f_x^0 = f_x^y = f_x^z = 0, \nonumber \\
&f_x^x = \frac{n_\mathrm{imp}u_0^2(a_0 - a_x + a_y - a_z)}{1 - n_\mathrm{imp}u_0^2(a_0 - a_x + a_y - a_z)},
\end{align}
where
\begin{align}
a_0(\omega) &= \int \frac{\omega^2}{R(\omega,\vec q)}\frac{d^3 q}{(2\pi)^3}, \ a_x(\omega) = \int \frac{(v_z\ell(q_x))^2}{R(\omega,\vec q)}\frac{d^3 q}{(2\pi)^3}, \nonumber \\
a_y(\omega) &= \int \frac{(v_yq_y)^2}{R(\omega,\vec q)}\frac{d^3 q}{(2\pi)^3}, \ a_z(\omega) = \int \frac{(v_zq_z)^2}{R(\omega,\vec q)}\frac{d^3 q}{(2\pi)^3}, \nonumber
\end{align}
with 
\begin{align}
	R(\omega,\vec q) \equiv K(\omega,\vec q)K^*(\omega,\vec q)\nonumber
\end{align}
and
\begin{align}
	K(\omega,\vec q) =&  \left(\omega + i\Gamma_{b,0}(\omega)\right)^2   \nonumber \\
	& - \left[(v_zq_z)^2 + (v_yq_y)^2 + (v_z\ell(q_x) + i\Gamma_{b,2}(\omega))^2\right]. \nonumber
\end{align}
It follows that vertex function $\Lambda_x$ is given by
\begin{align} \label{eq:Lambdax_lfstz}
\Lambda_{b,x} = \frac{1}{1 - n_\mathrm{imp}u_0^2(a_0 - a_x + a_y - a_z)}\sigma_x.
\end{align}

\begin{figure}
	\centering
	\subfigure[\label{fig:chib}]{\includegraphics[scale=0.3]{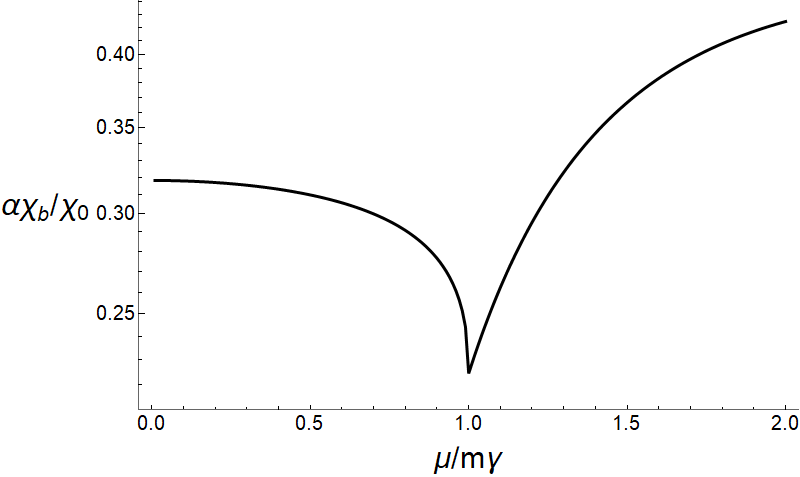}}
	\subfigure[\label{fig:chi}]{\includegraphics[scale=0.3]{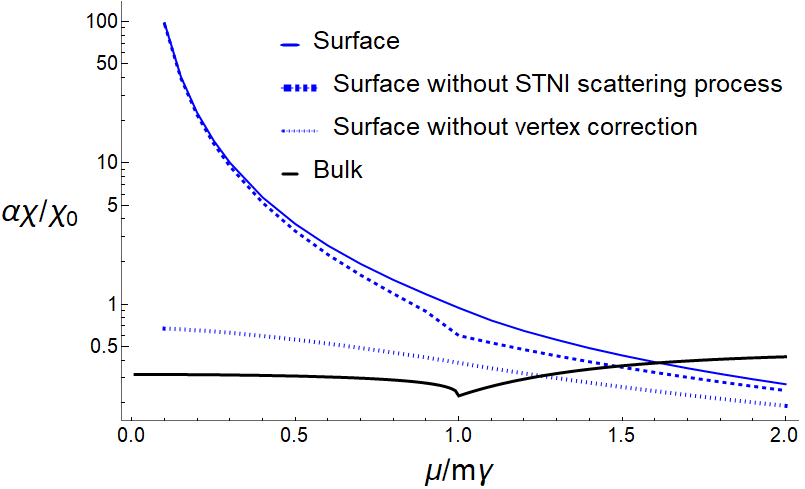}}
	\caption{(a) Plot of Bulk states' magnetoelectric susceptibility ($\chi_b$) against chemical potential ($\mu$) with $\alpha = 0.001$. We note that the shape of the plot is independent of $\alpha$ when $\alpha$ is small ($\alpha \lesssim 0.1$). (b) Comparison plots of magnetoelectric susceptibility vs. chemical potential for bulk states and various cases of surface states at $z = 0$. The parameters $\chi_0$ and $\alpha$ are defined in the caption of Fig. \ref{fig:chis}.}
\end{figure}

In order to evaluate the integrals in the $a_i$'s, we make a change of variables, $(v_z\ell(q_x),v_yq_y,v_zq_z) = |\varepsilon|\cos\theta,|\varepsilon|\sin\theta\cos\phi,|\varepsilon|\sin\theta\sin\phi)$. Under this change of variables, it turns out that $a_y(\omega) - a_z(\omega) = 0$. Furthermore, since $\frac{1}{R(\omega,\varepsilon,\theta)}$ is sharply peaked at $\varepsilon$ such that $\varepsilon^2 = \omega^2$, we make an approximation that $\varepsilon \approx \omega$ in the numerator (see \cite{Rodionov2015} for a thorough investigation of this approximation). Consequently, the expression of $a_0 - a_x + a_y - a_z = a_0 - a_x$ turns into 
\begin{align}
(a_0 - a_x)(\omega) = &\frac{\omega^4}{4\pi^2\gamma \sqrt{m}v_yv_z} \int d(\varepsilon,\theta) \nonumber \\ 
&\times \frac{\sin^3\theta}{R(\omega,\varepsilon,\theta)\sqrt{1-\frac{|\varepsilon|}{m\gamma}\cos\theta}},
\end{align}
where $\int d(\varepsilon,\theta) \equiv \int_0^{m\gamma}d\varepsilon \int_0^\pi\theta + \int_{m\gamma}^{\infty}d\varepsilon\int_{\cos^{-1}\left(\frac{m\gamma}{\varepsilon}\right)}^{\pi}d\theta$. Substituting $\Lambda_{b,x}$ from Eq. \ref{eq:Lambdax_lfstz} and the bulk Green function into Eq. \ref{eq:chib_formula}, we can simplify the expression for the bulk states' magnetoelectric susceptibility as
\begin{align}
\chi_{b} = \frac{2ev_y\mu_B}{\pi} \int_{-\infty}^{\infty}d\omega &\left(-\frac{\partial n_F(\omega - \mu)}{\partial \omega}\right) \nonumber \\
&\times \frac{(a_0 - a_x)(\omega)}{1-n_\mathrm{imp}u_0^2(a_0 - a_x)(\omega)}. \nonumber 
\end{align}
There is an additional factor of $2$ in this expression because the two pseudospin sectors equally contribute to the bulk response. As in the case of $\chi_s$, one can write $\chi_b$ in a scaling form (see Sec. \ref{app:dimless}) as
\begin{align} \label{eq:chib_scaling}
	\chi_b = \frac{2\chi_0}{\alpha} \tilde{\chi}_b\left(\frac{\mu}{m\gamma},\frac{T}{m\gamma},\alpha\right),
\end{align}
where $\tilde{\chi}_b$ is the dimensionless bulk susceptibility. From our numerical calculation, we find that $\tilde{\chi}_b$ is independent of $\alpha$ for $\alpha \lesssim 0.1$. This means, for small $\alpha$, the effect of impurity scatterings is captured entirely in a mulplication factor. We display the plot of $\chi_{b}$ vs. $\mu$ at $T = 0$ in Fig. \ref{fig:chib}. There is a dip at $\mu = m\gamma$ which correponds to Lifshitz transitions of the model. 

In Fig. \ref{fig:chi}, we overlay $\chi_s$ at $z = 0$ on top of the plot of $\chi_b$ to compare the surface and bulk results. We find that $\chi_s$ at the interface is much larger than $\chi_b$ at low chemical potentials. As $\mu$ increases, $\chi_s$ quickly decreases, whereas $\chi_b$ approximately stays in the range $0.2 - 0.4$ in units of $\chi_0/\alpha$. Eventually, $\chi_b$ becomes larger than $\chi_s$ once $\mu \gtrsim 1.6$. In order to investigate the origin of large surface response, we calculate $\chi_s$ in the absence of the STNI scattering process (i.e., $\Gamma_{STB}$ is set to equal $\Gamma_{STTI}$ in Eq. \ref{eq:sigma_b0_final}) and $\chi_s$ with no vertex corrections. The results are also plotted in Fig. \ref{fig:chi} to make a comparison with the full $\chi_s$. We find that when the STNI scattering process is excluded, only minimal value of $\chi_s$ decreases. However, when the vertex correction is neglected, there is a large drop in $\chi_s$ such that its value at $z=0$ has the same order of magnitude as $\chi_b$. 

\section{Discussion and Conclusion}

Let us try to understand the effects that a surface has on the Edelstein response. First, we consider two type of the surface-to-bulk scattering processes. The bulk Green functions in the presence of a surface have two contributions: the translationally invariant part, $G_{ti}$, which is simply the bulk Green function in the absence of any surface, and the nontranslationally invariant part, $G_{ni}$, which arises due to the correlation between the incident and the reflecting bulk wave functions. In the approximation in which the STNI process (or $G_{ni}$) is neglected, $\Gamma_{STB}$ comes entirely from $\Gamma_{STTI}$ as given by Eq. \ref{eq:sigma_b0_final}. We find that including $G_{ni}$ results in a smaller total $\Gamma_{STB}$ than $\Gamma_{STTI}$ as can be seen from Fig. \ref{fig:gammastb}. This change in the scattering rate indicates that the STNI process could affect the linear response of the surface states because the quasiparticles have a longer life-time. However, as shown in Fig. \ref{fig:chi}, the inclusion of the STNI process causes $\chi_s$ to increase slightly and, thus, only has a small effect on the Edelstein response. 

Second, we turn to the effect of the surface states' vertex correction. We can see from Eqs. \ref{eq:chi_s}, \ref{eq:XS_total}, and \ref{eq:XS_1} that, at fixed $\omega$, the vertex correction ($f(\omega,z_1,z_2)$) is directly related to $\chi_s$. We find from our calculation in Fig. \ref{fig:chi} that when $f(\omega,z_1,z_2)$ is neglected, $\chi_s$ at $z=0$ decreases by one to two orders of magnitude at low chemical potentials ($\mu \lesssim 0.5m\gamma$). This means $f(\omega,z_1,z_2)$ is large at small $\omega$ or $\mu$. Hence, we can identify the large surface vertex correction to be the main responsibility for the strong enhancement of Edelstein response close to the surface. We can understand how the large vertex correction at low chemical potentials comes about as follows. As the chemical potential gets closer to the Weyl points, the coupling  between the surface and bulk states are weaker and weaker. This can be seen from the fact that the surface-to-bulk scattering rate vanishes continuously as $\omega \rightarrow 0$. In this limit, the surface states which effectively behave like a chiral fermion in one dimension must be dissipationless. However, the conductivity of such a chiral fermion is finite if only the bare vertex is present \cite{Gorbar2016}. This means, in order for the conductivity to be very large or infinite, the vertex correction has to be included and much larger than the bare vertex. We note that this argument relies on the 1D chiral fermions having no dissipations in the present of a quenched disorder. It would be interesting to see how the results would change for other momentum relaxation mechanisms such as a long-range impurity or an electron-phonon interaction.

We next compare our results with the calculation of the Edelstein effect in Weyl semimetal TaAs from \cite{Johansson2018}. We note that Ref. \cite{Johansson2018} proposed the Hamiltonian of the form given in Eq. \ref{eq:weyl_hamiltonian} and applied it to TaAs which has 24 Weyl nodes in the Brillouin zone. In this paper, we focus solely on Eq. \ref{eq:weyl_hamiltonian}, which has 4 Weyl nodes. This means we do not consider the scattering between different pairs of Weyl nodes. We only include the scatterings within a pair. Ref. \cite{Johansson2018} computed a magnetic moment response due to an applied electric field within the semi-classical Boltzmann approach. They found that the surface states' magnetic moment is much stronger than the bulk states' magnetic moment near the surface (by about 1 to 2 orders of magnitude). The surface states' magnetic moment quickly decreases as one goes deeper into the bulk, whereas the bulk states' magnetic moment stays constant. This behavior is qualitatively similar to our result, $\chi_s(\mu)$, at the chemical potential $\mu \lesssim 0.5m\gamma$. However, Ref. \cite{Johansson2018} fixed the value of $\mu$ to be about $m\gamma$.\footnote{There are two types of Weyl nodes in TaAs, W1 and W2, but only Weyl nodes of the type W1 contribute to the Edelstein effect. Ref. \cite{Johansson2018} used the experimentally fitted parameters for W1, $\mu = 22.1 \ \mathrm{meV}$, $m = 1.73\times10^{-4} \mathrm{\AA}^{-2}$, and $\gamma = 130 \ \mathrm{eV}\mathrm{\AA}^2$, from \cite{Lee2015}. One finds the ratio $\mu/m\gamma = 0.982$.}  At that value of $\mu$, our $\chi_s$ is only about 4 times larger than $\chi_b$. This discrepancy could be attributed to the following possibilities. First, we do not include the scatterings between different pairs of Weyl nodes in the calculation. It is possible that including such processes could lead to much larger scattering rates for the bulk states than the surface states. The second possibility is that the Kubo formalism we use here could be capable of capturing electron-electron correlations that are not present in the semiclassical Boltzmann approach. Thus, our results may not necessarily have the same qualitative and quantitative features as \cite{Johansson2018}. Nonetheless, we show in this paper that it is sufficient for Weyl semimetals to have a strong surface Edelstein response without any scatterings between different Weyl-node pairs, albeit at lower chemical potentials.

To summarize, the key result of this paper is that, near the surface of Weyl semimetals, the surface states have a much stronger magnetoelectric susceptibility than the bulk states at low chemical potentials. The underlying reason for this phenomenon is that, at the chemical potential close to the Weyl points, the Fermi-arc surface states weakly couple to bulk states. The surface states become almost dissipationless and, thus, have a large vertex correction. Since the vertex corrections generally appear in the calculations of many correlation functions, we expect other responses and transport properties (e.g., thermal conductivity and thermopower) of Weyl semimetals to display a similar behavior.

\begin{acknowledgments}
We acknowledge Thailand Research Fund and Office of Higher Education Commission (Grant No. MRG6280130) and Faculty of Science, Mahasarakham University (Grant Year 2019) for funding of this project. We thank Chandan Setty for helpful comments on the manuscript.
\end{acknowledgments}

\onecolumngrid

\appendix

\section{Hermitian boundary conditions of the surface and bulk states} \label{app:hermitian_bc}
Some cares are needed when we pick the boundary conditions in order for the Hamiltonian on the left-hand side of Eq. \ref{eq:weyl_schrodinger} to be a Hermitian operator (see the discussion of this problem, for example, in \cite{Stone}). We want to pick the boundary conditions such that the hermicity condition,
\begin{align} \label{eq:hermitian}
\langle \psi | H |\xi \rangle = \langle  \xi | H | \psi \rangle^\dagger,
\end{align}
is satisfied for any states $| \psi \rangle$ and $| \xi \rangle$ in the Hilbert space. Since $k_{x}$ and $k_{y}$ are c-number, one needs to consider only the term with the operator $\frac{\partial}{\partial z}$. Using integration by parts on the left-hand side of Eq. \ref{eq:hermitian}, one finds
\begin{align}
	\int_{-\infty}^{\infty}dz \psi^\dagger \left(-i v_z \sigma_z \frac{\partial}{\partial z} \right) \xi  =  & -i v_z\psi^\dagger \sigma_z \xi \bigg\vert_{-\infty}^0 -i v_z\psi^\dagger \sigma_z \xi \bigg\vert_0^{\infty}  + \int_{-\infty}^{\infty}dz \left(-i v_z \sigma_z \frac{\partial}{\partial z} \psi\right)^\dagger  \xi.  
\end{align}
This means the Hamiltonian is Hermitian if the condition
\begin{align} \label{eq:hermitian_condition}
\psi^\dagger \sigma_z \xi \bigg\vert_{-\infty}^0 + \psi^\dagger \sigma_z \xi \bigg\vert_0^{\infty}  = 0
\end{align}
is satisfied.

In Sec. \ref{sec:surface}, we introduce an interface between Weyl semimetals and a vacuum at $z = 0$. The system is assumed to be Weyl semimetals for $z > 0$ and a vacuum for $z < 0$. Wave functions must vanish far inside the vacuum. This means $\psi^\dagger \sigma_z \xi(-\infty) = 0$. Furthermore, the continuity of wave functions at the interface means $\psi^\dagger \sigma_z \xi(0^+) = \psi^\dagger \sigma_z \xi(0^-)$. Therefore, the condition required for the Hamiltonian to be Hermitian is reduced from Eq. \ref{eq:hermitian_condition} to
\begin{align} \label{eq:hermitian_condition_reduced}
\psi^\dagger \sigma_z \xi (\infty) = 0.
\end{align}

In the case of surface states, since the wave functions are localized near the surface, they must vanish deep inside the bulk as $z \rightarrow \infty$. Hence, Eq. \ref{eq:hermitian_condition_reduced} is satisfied. The boundary conditions for the surface states can be summarized as
\begin{align}
\lim\limits_{z \rightarrow \pm\infty}\psi(z) &= 0 \label{eq:surfacebc1}  \\
\lim\limits_{z \rightarrow 0^+} \psi(z) &= \lim\limits_{z \rightarrow 0^-}\psi(z) \label{eq:surfacebc2}. 
\end{align}

In the case of bulk states, the wave functions do not simply vanish as $z\rightarrow\infty$. Let us expand Eq. \ref{eq:hermitian_condition_reduced} in terms of the components of the two wave functions, $\psi = \begin{pmatrix}
	\psi_1 \\
	\psi_2
\end{pmatrix}$ and $\xi = \begin{pmatrix}
\xi_1 \\
\xi_2
\end{pmatrix}$. One finds
\begin{align}
\left(\frac{\psi_1(\infty)}{\psi_2(\infty)}\right)^*  = \frac{\xi_2(\infty)}{\xi_1(\infty)}.
\end{align}
One possible solution of this equation is that the ratio of the components of wave functions at $z = \infty$ must be a complex phase, 
\begin{align} \label{eq:psi_ratio}
\frac{\psi_1(\infty)}{\psi_2(\infty)} = e^{i\phi}.
\end{align}
For the bulk states, we can interpret $z \rightarrow \infty$ as $z = L_z$ where $L_z$ can be thought of as the thickness of the Weyl semimetal along the $z$ direction. $L_z$ is a macroscopic length scale that is much larger than any other intrinsic length scales such as the inverse Fermi wave vector. From Eq. \ref{eq:psi_ratio}, we choose the boundary condition at $z = L_z$ to be $\frac{\psi_1(L_z)}{\psi_2(L_z)} = p = \pm 1$. The boundary conditions for the bulk states can be summarized as
\begin{align}
\lim\limits_{z \rightarrow -\infty}\psi(z) &= 0  \label{eq:bulkbc1} \\
\lim\limits_{z \rightarrow 0^+} \psi(z) &= \lim\limits_{z \rightarrow 0^-}\psi(z) \label{eq:bulkbc2}. \\
\frac{\psi_1(L_z)}{\psi_2(L_z)} &= p \label{eq:bulkbc3}.  
\end{align}
The choice that the ratio in Eq. \ref{eq:psi_ratio} equals $e^{i\phi} = p$ can be understood physically from a particular setup in which the region $0<z<L_z$ is a bulk of Weyl semimetal and the region above $z > L_z$ is a vacuum. That is there is an additional Weyl semimetal-vacuum interface located at $z = L_z$. Solving Eq. \ref{eq:weyl_schrodinger} with $m(z) = -\tilde{m}\rightarrow-\infty$ in the region $z > L_z$ yields an eigenstate wave function of a form
\begin{align}
\begin{pmatrix}
	\psi_1 (z)\\
	\psi_2 (z)
\end{pmatrix} = B \begin{pmatrix}
p \\
1
\end{pmatrix} e^{ik_x x + ik_y y - \frac{\gamma \tilde{m}}{v_z}(z - L_z)},
\end{align}
where $B$ is a constant. By invoking the continuity of the wave functions at the interface, we find that on the Weyl semimetal side $\frac{\psi_1(L_z^-)}{\psi_2(L_z^-)}= p$.

\section{Calculation of surface eigenstates} \label{app:surface}
The Schr\"odinger equation in Eq. \ref{eq:weyl_schrodinger} can be rewritten as
\begin{align} \label{eq:weyl_schrodinger_A}
-i \frac{\partial}{\partial z} \psi + A\psi = 0
\end{align}
where the matrix $A$ is given by
\begin{align} \label{eq:matrixA}
A = \begin{pmatrix}
	-\varepsilon/v_z & -\frac{v_y}{v_z}k_{y} + i p\ell(k_x) \\
	\frac{v_y}{v_z}k_{y} + i p\ell(k_x) & \varepsilon/v_z
\end{pmatrix}.
\end{align}
Here, we define $\ell(k_x) \equiv \frac{\gamma}{v_z}(m-k_x^2)$. Solving the characteristic equation, $|A - \lambda I| = 0$, one finds,
\begin{align} \label{eq:lambda}
\lambda^2 = \left( \frac{\varepsilon}{v_z} \right)^2 -  \ell^2(k_x)  -  \left( \frac{v_y}{v_z}k_{y} \right)^2.
\end{align}
Depending on the phase of the system, $\lambda$ can be real or pure imaginary. In the case of surface states, the wave functions are localized near the surface and, thus, must vanish as $z\rightarrow \pm \infty$ (see Eq. \ref{eq:surfacebc1}). This means $\lambda$ must be pure imaginary (otherwise the wave function would be extended). We can write down $\lambda$ as
\begin{align}  \label{eq:eigenvalues_surface}
\lambda^s_{\pm} = \mp i\kappa,
\end{align}
where
\begin{align} \label{eq:kappa}
\kappa = \sqrt{\ell^2(k_x) + \left(\left(\frac{v_y}{v_z}\right) k_{y}\right)^2 - \left(\frac{\varepsilon}{v_z}\right)^2 }.
\end{align}
The eigenvectors associated with the eigenvalues $\lambda_{s,\pm}$ are
\begin{align} \label{eq:eigenvectors_surface}
\begin{pmatrix} 
	v_1 \\ v_2
\end{pmatrix}_{s,\pm}
= 
\begin{pmatrix}
	\frac{v_y}{v_z}k_{y} - i p\ell(k_x) \\
	-\varepsilon/v_z\pm i \kappa
\end{pmatrix}.
\end{align}
The general solution of the surface wave functions is
\begin{align}
\begin{pmatrix} 
	\psi_1 \\ \psi_2
\end{pmatrix}_{s} = & B_+ 
\begin{pmatrix} 
	v_1 \\ v_2
\end{pmatrix}_{s,+} e^{-\kappa z + ik_{x}x + ik_{y}y} 
 + B_-
\begin{pmatrix} 
	v_1 \\ v_2
\end{pmatrix}_{s,-} e^{\kappa z + ik_{x}x + ik_{y}y}.
\end{align}

Using Eq. \ref{eq:surfacebc2}, we have $A_+ = 0$ for $z<0$ and $A_- = 0$ for $z>0$. That is
\begin{align} \label{eq:eigenstates_surface_1}
\begin{pmatrix} 
	\psi_1 \\ \psi_2
\end{pmatrix}_{s} = \begin{cases}
	B_+ 
	\begin{pmatrix} 
		v_1 \\ v_2
	\end{pmatrix}_{s,+} e^{-\kappa^+ z + ik_{x}x + ik_{y}y}  & \mathrm{for} \ z>0 \\
	B_-
	\begin{pmatrix} 
		v_1 \\ v_2
	\end{pmatrix}_{s,-} e^{\kappa^- z + ik_{x}x + ik_{y}y} & \mathrm{for} \ z<0
\end{cases}. 
\end{align} 
Here the superscript $+$ and $-$ denotes the values of $\kappa$ from the regions $z>0$ and $z<0$, respectively. Using the continuity of wave functions at the interface (Eq. \ref{eq:surfacebc1}), one has
\begin{align}
B_+ \begin{pmatrix} 
	v_1 \\ v_2
\end{pmatrix}_{s,+} 
= B_-\begin{pmatrix} 
	v_1 \\ v_2
\end{pmatrix}_{s,-} \nonumber 
\end{align}
Solving this equation using Eq. \ref{eq:eigenvectors_surface} and the definition of $\kappa^{\pm}$ from Eq. \ref{eq:kappa}, we find the energy of the surface eigenstates as
\begin{align}
\varepsilon^s_k = p v_y k_{y}.
\end{align}
Plugging Eq. \ref{eq:surface_states_energy} into Eq. \ref{eq:kappa}, we have
\begin{align}
\kappa = \frac{\gamma}{v_z}|k_{x}^2 - m| = 
\begin{cases} \label{eq:kappa2}
	\ell(k_x) & \mathrm{for} \ z>0 \\
	\frac{\gamma}{v_z}\tilde{m} & \mathrm{for} \  z<0.
\end{cases}
\end{align}
Using Eqs. \ref{eq:surface_states_energy} and \ref{eq:kappa2}, the eigenstates from Eq. \ref{eq:eigenvectors_surface} reduce to
\begin{align}
\begin{pmatrix} 
	v_1 \\ v_2
\end{pmatrix}_{s,\pm}
\propto
\begin{pmatrix}
	-p \\
	1
\end{pmatrix}
\end{align}
in the limit $\tilde{m}\rightarrow\infty$.

Hence, the eigenstate from Eq. \ref{eq:eigenstates_surface_1} is
\begin{align} \label{eq:eigenstates_surface_2}
\begin{pmatrix} 
	\psi_1 \\ \psi_2
\end{pmatrix}_{s} = \begin{cases}
	\frac{1}{\sqrt{N}} 
	\begin{pmatrix} 
		-p \\ 1
	\end{pmatrix} e^{-\ell(k_x) z + ik_{x}x + ik_{y}y}  & \mathrm{for} \ z>0 \\
	\frac{1}{\sqrt{N}}
	\begin{pmatrix} 
		-p \\ 1
	\end{pmatrix} e^{\frac{\gamma\tilde{m}}{v_z} z + ik_{x}x + ik_{y}y}  & \mathrm{for} \  z<0
\end{cases}. 
\end{align}
Here, we use the continuity of wave functions at the interface (Eq. \ref{eq:surfacebc2}) to show that the two coefficients in front must be equal $B_+ = B_- \equiv \frac{1}{\sqrt{N}}$. The constant $N$ can be determined from the normalization condition,
\begin{align}
\int^{L/2}_{-L/2} dx \int^{L/2}_{-L/2} dy \int_{0}^{\infty} dz 
\psi_s^\dagger \psi_s= 1,
\end{align}
where we let $\begin{pmatrix}\psi_1 \\ \psi_2\end{pmatrix} \rightarrow 0$ for $z<0$ by taking the limit $\tilde{m}\rightarrow0$. Finally, the surface eigenstates (for $z>0$) are
\begin{align}
\begin{pmatrix} 
	\psi_1 \\ \psi_2
\end{pmatrix}_{s} = 
\frac{\sqrt{\ell(k_x)}}{L}
\begin{pmatrix} 
	-p \\ 1
\end{pmatrix} e^{-\ell(k_x)z + ik_{x}x + ik_{y}y}.
\end{align}

\section{Calculation of bulk eigenstates} \label{app:bulk}

In the case of bulk eigenstates, the wave functions are localized in the vacuum and extended inside the materials. This is to be contrast with the surface eigenstates in which the wave functions are localized both in the vacuum and inside the Weyl semimetals. In the vacuum region ($z<0$), one can immediately write down the wave function in the vacuum from Eq. \ref{eq:eigenstates_surface_2} as
\begin{align} \label{eq:vacuum_wf}
\begin{pmatrix}
	\psi_1 \\ \psi_2
\end{pmatrix}
= 
B\begin{pmatrix}
	-p \\
	1 
\end{pmatrix}  e^{ik_{x}x + ik_{y}y+\frac{\gamma \tilde{m}}{v_z} z},
\end{align}
where $B$ is a constant. In the Weyl semimetal-region $0<z<L_z$, the eigenvalues are real. Hence, form Eq. \ref{eq:lambda}, we have $\lambda_\pm = \mp\tilde{k}_z$, where $\tilde{k}_z = \sqrt{\left( \frac{\varepsilon}{v_z} \right)^2 - \left( v_z\ell(k_x) \right)^2  -  \left(\frac{v_y}{v_z}k_{y} \right)^2}$. One can rewrite the energy $\varepsilon$ in term of $\tilde{k}_z$ as
\begin{align} \label{eq:eigenenergy_bulk}
\varepsilon = \pm\sqrt{(v_z \ell(k_x))^2 + (v_y k_{y})^2 + ( v_z \tilde{k}_z)^2}.
\end{align}
In the absence of surface, $\tilde{k}_z$ is replaced by the momentum along the z direction $k_{z}$ and $\varepsilon$ becomes the bulk energy spectrum without any interface. The corresponding eigenvectors to matrix $A$ (Eq. \ref{eq:matrixA}) are
\begin{align}
\begin{pmatrix}
	v_1 \\ v_2
\end{pmatrix}_{b,\pm}
= 
\begin{pmatrix}
	v_y k_{y} - i  pv_z\ell(k_x) \\
	-\varepsilon \pm v_z \tilde{k}_z 
\end{pmatrix} \nonumber.
\end{align}
The general solution in the region $0<z<L_z$ is
\begin{align} \label{eq:weylslab_gen_sol} 
\begin{pmatrix}
	\psi_1 \\ \psi_2
\end{pmatrix}_b
=&
C_+\begin{pmatrix}
	v_1 \\ v_2
\end{pmatrix}_{b,+} e^{ik_{x}x + ik_{y}y - i\lambda_+ z} 
 +C_-\begin{pmatrix}
	v_1 \\ v_2
\end{pmatrix}_{b,-} e^{ik_{x}x + ik_{y}y - i\lambda_- z}. 
\end{align}
Using the continuity of wave function at the interface $z = 0$ (Eq. \ref{eq:bulkbc2}) with Eqs. \ref{eq:vacuum_wf} and \ref{eq:weylslab_gen_sol} yields two equations. We can get rid of the constant $B$ (from Eq. \ref{eq:vacuum_wf}) by taking the ratio of these two equations. With a few steps of algebra, one obtains
\begin{align} \label{eq:C+C-bc0}
\frac{C_+}{C_-} = \frac{-v_y k_{y} + i  pv_z\ell(k_x) +p\varepsilon + pv_z \tilde{k}_z}{v_y k_{y} - i  pv_z\ell(k_x) - p\varepsilon + pv_z \tilde{k}_z}
\end{align}
Using the boundary condition at $z = L_z$ (Eq. \ref{eq:bulkbc3}) with Eq. \ref{eq:weylslab_gen_sol} and following the same manipulation as Eq. \ref{eq:C+C-bc0} result in
\begin{align} \label{eq:C+C-bcLz}
\frac{C_+}{C_-}e^{2i\tilde{k}_z L_z} = \frac{-v_y k_{y} + i  pv_z\ell(k_x) - p\varepsilon - p v_z \tilde{k}_z}{v_y k_{y} - i  pv_z\ell(k_x) + p\varepsilon - p v_z \tilde{k}_z}.
\end{align}
Dividing Eq. \ref{eq:C+C-bcLz} by Eq. \ref{eq:C+C-bc0}, substituting in the energy $\varepsilon$ from Eq. \ref{eq:eigenenergy_bulk}, and simplifying the expression yield
\begin{align}
e^{2i\tilde{k}_z L_z} = -\frac{\tilde{k}_z - i \ell(k_x)}{\tilde{k}_z + i \ell(k_x)} = e^{2i\phi + (2n + 1)\pi},
\end{align}
where $\tan\phi = -\frac{\ell(k_x)}{\tilde{k}_z}$ and $n \in \mathbb{Z}$ such that $\tilde{k}_z > 0$. It follows that 
\begin{align}
\tilde{k}_z = \frac{\phi}{L_z} + \left(n+\frac{1}{2}\right)\frac{\pi}{L_z}.
\end{align}
In the limit $L_z \rightarrow \infty$, the phase term $\frac{\phi}{L_z} \rightarrow 0$ whereas the term $\left(n + \frac{1}{2}\right)\frac{\pi}{L_z}$ can be finite if $n$ is sufficiently large (in such a way that $n/L_z$ is finite). Furthermore, the distant between adjacent $k_z$ is $\Delta k_z = \pi/L_z \rightarrow 0$. Hence, $\tilde{k}_z$ becomes strictly positive and continuous as $L_z \rightarrow \infty$.

From Eq. \ref{eq:C+C-bc0}, we write the constant $C_+$ and $C_-$ in term of a single constant $C$ as
\begin{align}
C_+ &= \left(-v_y k_{y} + i  pv_z\ell(k_x)+p\varepsilon + p v_z \tilde{k}_z \right)C, \nonumber \\
C_- &= \left(v_y k_{y} - i  pv_z\ell(k_x)-p\varepsilon + p v_z \tilde{k}_z \right)C \nonumber.
\end{align}
The solution in the region $0<z<L_z$ becomes
\begin{align}  \label{eq:weylslab_gen_sol_2} 
\begin{pmatrix}
	\psi_1 \\ \psi_2
\end{pmatrix}
=&
C\bigg[ \left(-v_y k_{y} + i  pv_z\ell(k_x) + p\varepsilon + p v_z \tilde{k}_z \right)
\begin{pmatrix}
	v_y k_{y} - i  pv_z\ell(k_x) \\
	-\varepsilon + v_z \tilde{k}_z 
\end{pmatrix}  e^{ik_{x}x + ik_{y}y + i\tilde{k}_z z} \nonumber \\
& \ \ +
\left(v_y k_{y} - i  pv_z\ell(k_x) - p\varepsilon + p v_z \tilde{k}_z \right)
\begin{pmatrix}
	v_y k_{y} - i  pv_z\ell(k_x) \\
	-\varepsilon - v_z \tilde{k}_z 
\end{pmatrix}  e^{ik_{x}x + ik_{y}y - i\tilde{k}_z z} \bigg]. 
\end{align}
The constant $C$ can be determined by normalization condition,
\begin{align}
\int^{L/2}_{-L/2} dx \int^{L/2}_{-L/2} dy \int_{0}^{L_z} dz \psi_b^\dagger \psi_b = 1.
\end{align}
Performing the normalization integral results in
\begin{align}
1 = |C|^2 L^2
	\bigg\{ & L_z \left( (p\varepsilon + p v_z \tilde{k}_z - v_y k_{y})^2 + (v_z\ell(k_x))^2 \right)\left( (v_y k_{y})^2 + (v_z\ell(k_x))^2 + (-\varepsilon + v_z \tilde{k}_z)^2 \right) \nonumber \\
	+ & L_z \left( (-p\varepsilon + p v_z \tilde{k}_z + v_y k_{y})^2 + (v_z\ell(k_x))^2 \right)\left( (v_y k_{y})^2 + (v_z\ell(k_x))^2 + (\varepsilon + v_z \tilde{k}_z)^2 \right) \nonumber \\
	+ & \frac{e^{2i\tilde{k}_z L_z} - 1}{2i\tilde{k}_z} F(k_{x},k_{y},\tilde{k}_z) + \frac{e^{-2i\tilde{k}_z L_z} - 1}{-2i\tilde{k}_z} G(k_{x},k_{y},\tilde{k}_z) \nonumber \bigg\},
\end{align}
where the functions $F$ and $G$ come from the cross terms in the product $\psi_b^\dagger\psi_b$. We note that the first two terms are of the order $O(L_z)$ whereas the last two are of the order $O(1)$. This means one can neglect the last two terms in the limit $L_z \rightarrow \infty$. The constant $C$ can be simplified to 
\begin{align}
C = \frac{1}{\sqrt{8V_w \varepsilon (\varepsilon - p v_y k_{y})(\varepsilon^2 - (v_z \tilde{k}_z)^2)}},
\end{align}
where $V_w \equiv L^2 L_z$ is the volume of the Weyl semimetals. Finally, the bulk eigenstates in the presence of a surface located at $z = 0$ is given by
\begin{align}
\psi_b(\vec r)
=
A(\vec k_{\|},\tilde{k}_z)  e^{i(\vec k_{\|}\cdot\vec r_{\|} + \tilde{k}_z z)}  - A(\vec k_{\|},-\tilde{k}_z) e^{i(\vec k_{\|}\cdot\vec r_{\|} - \tilde{k}_z z)},
\end{align}
where 
\begin{align}
A(\vec k_{\|},\tilde{k}_z) \equiv & \frac{ - v_y k_{y} + i  pv_z \ell(k_x) +p\varepsilon + p v_z \tilde{k}_z }{\sqrt{8V_w \varepsilon (\varepsilon - p v_y k_{y})(\varepsilon^2 - (v_z \tilde{k}_z)^2)}} 
\begin{pmatrix}
	v_y k_{y} - i  pv_z\ell(k_x) \\
	-\varepsilon +  v_z \tilde{k}_z
\end{pmatrix}.
\end{align}
Here, $\vec k_{\|} \equiv (k_x, k_y)$ and $\vec r_{\|} \equiv (x,y)$ are the momentum and position vector parallel to the surface, respectively.

\section{Calculation of the bulk Green function in the presence of surface}
The bulk Green function can be computed from eigenstates using Eq. \ref{eq:green_func_eigenstates} with the sum over all bulk eigenstates from Eq. \ref{eq:psi_b}. The matrix $\psi_{b,\varepsilon}(\vec r) \psi_{b,\varepsilon}^\dagger(\vec r')$ can be calculated as
\begin{align} \label{eq:psipsidagger}
	\psi_{b,\varepsilon}(\vec r) \psi_{b,\varepsilon}^\dagger(\vec r') = \sum_{i = 0}^{1}\left[M_i(\vec k_{\|},\tilde{k}_z,\varepsilon) + M_i(\vec k_{\|},-\tilde{k}_z,\varepsilon)\right]
\end{align}
where $M_0$ and $M_1$ are matrices of the form
\begin{align}
	M_0(\varepsilon,k) =&  A(\vec k_{\|},\tilde{k}_z)A^\dagger(\vec k_{\|},\tilde{k}_z)e^{i(\vec k_{\|}\cdot(\vec r_{\|} - \vec r'_{\|}) + \tilde{k}_z(z-z'))} \nonumber \\
	= &
	\frac{1}{4V_w \varepsilon}\begin{pmatrix}
		\varepsilon +  v_z \tilde{k}_z & - v_y k_{y} + ipv_z\ell(k_x) \\
		- v_y k_{y} - ipv_z\ell(k_x) & \varepsilon -  v_z \tilde{k}_z
	\end{pmatrix}   e^{i(\vec k_{\|}\cdot(\vec r_{\|} - \vec r'_{\|}) + i\tilde{k}_z (z-z'))} 
\end{align} 
and
\begin{align}
	M_1(\varepsilon,k) = &  -A(\vec k_{\|},\tilde{k}_z)A^\dagger(\vec k_{\|},-\tilde{k}_z)e^{i(\vec k_{\|}\cdot(\vec r_{\|} - \vec r'_{\|}) + \tilde{k}_z(z+z'))} \nonumber \\
	=& \begin{pmatrix}
		\varepsilon^2 - ( v_z \tilde{k}_z)^2 & -(\varepsilon +  v_z \tilde{k}_z)( v_y k_{y} - ipv_z\ell(k_x)) \\
		-(\varepsilon -  v_z \tilde{k}_z)( v_y k_{y} + ipv_z\ell(k_x)) & \varepsilon^2 - ( v_z \tilde{k}_z)^2
	\end{pmatrix} \nonumber \\
	&\times \frac{p v_y k_{y}(\varepsilon-p v_y k_{y})+v_z\ell(k_x)(-v_z\ell(k_x)+i v_z \tilde{k}_z)}{4V_w \varepsilon (\varepsilon - p v_y k_{p,y})(\varepsilon^2 - ( v_z \tilde{k}_z)^2)} e^{i(\vec k_{\|}\cdot(\vec r_{\|} - \vec r'_{\|}) + \tilde{k}_z(z+z'))} 
\end{align} 

The summation over all eigenstates in Eq. \ref{eq:green_func_eigenstates} can be converted into integrals over $(k_x,k_y,\tilde{k}_z)$ and the sum of positive- and negative-energy states as $\sum_{E}\frac{\psi_{E} \psi_{E}^\dagger}{\omega - E} =  2V_w\int \frac{d^2 k_{\parallel}}{(2\pi)^2}\int_{0}^{\infty}\frac{d\tilde{k}_z}{2\pi} \left( \frac{\psi_{\varepsilon_k} \psi_{\varepsilon_k}^\dagger}{\omega - \varepsilon_k} + \frac{\psi_{-\varepsilon_k} \psi_{-\varepsilon_k}^\dagger}{\omega + \varepsilon_k} \right)$. We note that, since $\tilde{k}_z \approx \frac{(n+1/2)\pi}{L_z}>0$, the summation over $\tilde{k}_z$ is converted into integral as $\sum_{\tilde{k}_z} \rightarrow 2L_z\int_{0}^{\infty}\frac{d\tilde{k}_z}{2\pi}$. Substituting Eq. \ref{eq:psipsidagger} into Eq. \ref{eq:green_func_eigenstates}, one obtains
\begin{align}
G_b(\omega,\vec r_{\parallel}-\vec r'_{\parallel},z,z') &= 2V_w\int \frac{d^2 k_{\parallel}}{(2\pi)^2}\int_{0}^{\infty}\frac{d\tilde{k}_z}{2\pi}  \sum_{i=0}^{1}\left(\frac{M_i(\vec k_{\|},\tilde{k}_z,\varepsilon_k) + M_i(\vec k_{\|},-\tilde{k}_z,\varepsilon_k)}{\omega - \varepsilon_k} + \frac{M_i(\vec k_{\|},\tilde{k}_z,-\varepsilon_k) + M_i(\vec k_{\|},-\tilde{k}_z,-\varepsilon_k)}{\omega + \varepsilon_k} \right)\nonumber \\
&= 2V_w\int \frac{d^2 k_{\parallel}}{(2\pi)^2}\int_{-\infty}^{\infty}\frac{d\tilde{k}_z}{2\pi}  \sum_{i=0}^{1}\left(\frac{M_i(\vec k_{\|},\tilde{k}_z,\varepsilon_k)}{\omega - \varepsilon_k} + \frac{M_i(\vec k_{\|},\tilde{k}_z,-\varepsilon_k) }{\omega + \varepsilon_k} \right). \label{eq:GbMi}
\end{align}
On the second line, by combining the functions that depend on $\tilde{k}_z$ and $-\tilde{k}_z$, the limit of the integral over $\tilde{k_z}$ turns into $(-\infty,\infty)$.
Let us compute the contribution to the Green function from $M_0$. Substituting $M_0$ into Eq. \ref{eq:GbMi} and performing Fourier transform over the parallel coordinates yield 
\begin{align}
	G_{ti}(\omega,\vec k_{\parallel},&z,z') = \int_{-\infty}^{\infty} \frac{d\tilde{k}_z}{2\pi} \frac{e^{i\tilde{k}_z(z - z')}}{\omega^2 - \varepsilon_k^2}  \left( \omega \mathbbm{1} +  v_z \tilde{k}_z \sigma_z -  v_y k_{y} \sigma_x - pv_z\ell(k_x)\sigma_y \right)
\end{align}

In a similar manner to the calculation of $G_{ti}$, substituting $M_1$ into Eq. \ref{eq:GbMi} and simplifying the expression, we obtain the contribution to the bulk Green function from $M_1$ as
\begin{align}
	G_{ni}(\omega,\vec k_{\parallel},&z,z') = \int_{-\infty}^{\infty} \frac{d\tilde{k}_z}{2\pi} e^{i\tilde{k}_z(z + z')}\frac{1}{(\omega^2 - \varepsilon_k^2)}  \left[ \frac{p v_y \tilde{k}_z k_{y} - i\ell(k_x)\omega}{\tilde{k}_z - i\ell(k_x)}\mathbbm{1} - \frac{p \tilde{k}_z \omega + i v_y k_{y}\ell(k_x)}{ \tilde{k}_z - i\ell(k_x)}\sigma_x  - ipv_z(\tilde{k}_z + i\ell(k_x))\sigma_y \right]
\end{align}

The total bulk Green function in the presence of surface,
\begin{align}
G_{b}(\omega,\vec k_{\parallel},z,z') = G_{ti}(\omega,\vec k_{\parallel},z,z')+G_{ni}(\omega,\vec k_{\parallel},z,z'), 
\end{align}
is a sum of a function which is translationally symmetric ($G_{ti}$) and a function which breaks a translational symmetry ($G_{ni}$). 

\section{Calculations of the surface self-energies from various scattering processes} \label{app:sigma_s}
The calculations of the surface self-energies from the surface-to-surface and the surface-to-bulk processes were previously studied by Ref. \cite{Gorbar2016}. Since these self-energies are used in Secs. \ref{sec:surface_selfen} and \ref{sec:surface_response}, we briefly review how to calculate them in this appendix. From Eq. \ref{eq:sigma_s_2s}, one makes a replacement $\omega \rightarrow \omega + i\eta$ in the integrand to covert the self-energy into a retarded response. The surface self-energy from the surface-to-surface scattering can then be computed as
\begin{align}
\Sigma_{STS}
=& n_\mathrm{imp}u_0^2\int \frac{d^2q_{\|}}{(2\pi)^2}\frac{1}{\omega+i\eta-pv_yq_{y}} \frac{2\ell(k_{x})\ell(q_{x})}{\ell(k_{x})+\ell(q_{x})} \nonumber \\
=& -i\frac{n_\mathrm{imp} u_0^2}{2\pi v_y}\ell(k_{x})\int_{-\sqrt{m}}^{\sqrt{m}}dq_{x} \left(1 - \frac{\ell(k_{x})}{\ell(k_{x}) + \ell(q_{x})}  \right) \nonumber \\
=& -i \frac{n_\mathrm{imp}u_0^2}{\pi v_y }\ell(k_x) \left[\sqrt{m}-\frac{m-k_{x}^2}{\sqrt{2m-k_{x}^2}}\tanh^{-1}\left(\frac{\sqrt{m}}{\sqrt{2m-k_{x}^2}}\right) \right]. \nonumber 
\end{align}
On the first line, the integral over $q_{y}$ can be performed as
\begin{align}
&\int_{-\infty}^{\infty} \frac{dq_{y}}{\omega - pv_y q_{y} + i\eta} \nonumber \\
&= \int_{-\infty}^{\infty}dq_{y} \bigg(P\frac{1}{\omega - pv_y q_{y}} - i\pi\delta(\omega - pv_yq_{y}) \bigg) \nonumber \\
&= -i\frac{\pi}{v_y}. \nonumber
\end{align}
On the second line, we integrate over $q_{x}$ as
\begin{align}
\int_{-\sqrt{m}}^{\sqrt{m}}\frac{\ell(k_x)dq_{x}}{\ell(k_{x})+\ell(q_{x})} = (m-k_x^2)\int_{-\sqrt{m}}^{\sqrt{m}}dq_x \frac{1}{2m-q_x^2 - k_x^2} \nonumber \\
= \frac{2(m-k_x^2)}{\sqrt{2m-k_{x}^2}}\tanh^{-1}\left(\frac{\sqrt{m}}{\sqrt{2m-k_{x}^2}}\right). \nonumber
\end{align}

Let us turn to the surface-to-bulk process. From Eq. \ref{eq:sigma_b0_a}, one can calculate the imaginary part of $\Sigma_{STTI}$ from
\begin{align}
\mathrm{Im} \Sigma_{STTI} = -\pi n_\mathrm{imp}u_0^2 |\omega| \int\frac{d^3q}{(2\pi)^3}\delta(\omega^2 - \varepsilon_q^2) \nonumber
\end{align}
Substituting the energy spectrum, $\varepsilon_q$, and making a change of variables, $\vec Q = (Q_x,Q_y) = (\frac{v_y}{v_z}q_{y},\tilde{q}_z)$, we find
\begin{align}
\mathrm{Im} \Sigma_{STTI} =& -\frac{n_\mathrm{imp}u_0^2 |\omega|v_z}{8\pi^2 v_y} \int_{-\infty}^\infty dq_{x} \int_{0}^{\infty}dQ  2\pi Q \delta(\omega^2 - v_z^2 Q^2 - v_x^2 q_x^2) \nonumber
\end{align}
Performing an integral over $Q$ using the delta function results in
\begin{align} \label{eq:imsig_sb0_int_qx}
\mathrm{Im} \Sigma_{STTI} = -\frac{n_\mathrm{imp}u_0^2 |\omega|}{8\pi v_y v_z} \int_{-\infty}^\infty dq_{x} \theta\left(\frac{\omega^2}{v_z^2} - \ell^2(q_{x})\right)
\end{align}
The theta function means that $\frac{\omega^2}{v_z^2} - \ell^2(q_{x} > 0)$. Solving this inequality, we find the ranges of possible values of $q_x$ as follows.
In the case $\frac{|\omega|}{\gamma} > m$, the values of $q_{x}$ are 
\begin{align} \label{eq:qxrange1}
-\sqrt{m + \frac{|\omega|}{\gamma}} < q_{x} < \sqrt{m + \frac{|\omega|}{\gamma}}.
\end{align}
On the other hand, if $\frac{|\omega|}{\gamma} < m$, the values of $q_{x}$ are 
\begin{align} \label{eq:qxrange2}
-\sqrt{m + \frac{|\omega|}{\gamma}} < q_{x} < -\sqrt{m - \frac{|\omega|}{\gamma}} \ \ \mathrm{or} \ \ \sqrt{m - \frac{|\omega|}{\gamma}} < q_{x} < \sqrt{m + \frac{|\omega|}{\gamma}}.
\end{align}
By combining the two cases, we can perform the integral in Eq. \ref{eq:imsig_sb0_int_qx},
\begin{align}
\int_{-\infty}^\infty dq_{x} \theta\left(\frac{\omega^2}{v_z^2} - \ell^2(q_{x})\right) &= \theta\left(-m + \frac{|\omega|}{\gamma}\right)\int_{-\sqrt{m + \frac{|\omega|}{\gamma}}}^{\sqrt{m + \frac{|\omega|}{\gamma}}}dq_{x} + \theta\left(m - \frac{|\omega|}{\gamma}\right)\left(\int_{-\sqrt{m + \frac{|\omega|}{\gamma}}}^{-\sqrt{m - \frac{|\omega|}{\gamma}}}dq_{x} + \int_{\sqrt{m - \frac{|\omega|}{\gamma}}}^{\sqrt{m + \frac{|\omega|}{\gamma}}}dq_{x}\right) \nonumber \\
&= 2\left( \sqrt{m + \frac{|\omega|}{\gamma}} -\theta\left(m - \frac{|\omega|}{\gamma}\right)\sqrt{m - \frac{|\omega|}{\gamma}} \right).
\end{align}
Finally, substituting the result back into Eq. \ref{eq:imsig_sb0_int_qx}, we have 
\begin{align}
\mathrm{Im} \Sigma_{STTI} =& -\frac{n_\mathrm{imp}u_0^2 |\omega|}{4\pi v_y v_z} \left( \sqrt{m + \frac{|\omega|}{\gamma}} -\theta\left(m - \frac{|\omega|}{\gamma}\right)\sqrt{m - \frac{|\omega|}{\gamma}} \right). \nonumber
\end{align}

\section{Calculation of the bulk scattering rate $\Gamma_{b,2}$} \label{app:tau1}
In this appendix, we discuss how to calculate the bulk scattering rate $\Gamma_{b,2}$. As shown in Eq. \ref{eq:Gamma_b2}, $\Gamma_{b,2}$ is given by
\begin{align}
\Gamma_{b,2}(\omega) = -\frac{n_\mathrm{imp}u_0^2 v_z \mathrm{sgn}(\omega)}{8\pi^2}\int \delta(\omega^2 - \varepsilon_q^2)\ell(q_x) d^3q. \nonumber
\end{align}
The integral over $\vec q$ can be evaluated using the same technique as the calculation for $\mathrm{Im}\Sigma_{STTI}$ in Appendix \ref{app:sigma_s}. Making a change of variable $\vec Q = (Q_x,Q_y) = (\frac{v_y}{v_z}q_{y},\tilde{q}_z)$ and then integrating over $Q$, we have
\begin{align}	
\int \delta(\omega^2 - \varepsilon_q^2)\ell(q_x) d^3q = \frac{\pi}{v_y v_z}\int_{-\infty}^{\infty}dq_x \ell(q_x)\theta\left(\frac{\omega^2}{v_z^2} - \ell^2(q_x)\right). \nonumber
\end{align}
The Heaviside theta function means that the integral over $q_x$ has the limits of integration as in Eqs. \ref{eq:qxrange1} and \ref{eq:qxrange2}, 
\begin{align}
\int_{-\infty}^{\infty}dq_x \ell(q_x)\theta\left(\frac{\omega^2}{v_z^2} - \ell^2(q_x)\right)  &= \theta\left(-m+\frac{|\omega|}{\gamma}\right)\int_{-\sqrt{m+\frac{|\omega|}{\gamma}}}^{\sqrt{m+\frac{|\omega|}{\gamma}}} \ell(q_x) dq_x  + \theta\left(m - \frac{|\omega|}{\gamma}\right)\left( \int_{-\sqrt{m+\frac{|\omega|}{\gamma}}}^{-\sqrt{m-\frac{|\omega|}{\gamma}}} + \int_{\sqrt{m-\frac{|\omega|}{\gamma}}}^{\sqrt{m+\frac{|\omega|}{\gamma}}}  \right) \ell(q_x)dq_x \nonumber \\
&=  2\left[ m\sqrt{m+\frac{|\omega|}{\gamma}} - \frac{1}{3}\left(m+\frac{|\omega|}{\gamma}\right)^{\frac{3}{2}} - \theta\left(m - \frac{|\omega|}{\gamma}\right)\left( m\sqrt{m - \frac{|\omega|}{\gamma}} - \frac{1}{3}\left(m - \frac{|\omega|}{\gamma}\right)^{\frac{3}{2}} \right) \right]\nonumber
\end{align}
Thus, the bulk scattering rate $\Gamma_{b,2}$ is calculated to be
\begin{align}
&\Gamma_{b,2}(\omega) = -\frac{n_\mathrm{imp}u_0^2\gamma \mathrm{sgn}(\omega)}{4\pi v_y v_z} \left\{ m\sqrt{m+\frac{|\omega|}{\gamma}} - \frac{1}{3}\left(m+\frac{|\omega|}{\gamma}\right)^{\frac{3}{2}}  - \theta\left(m - \frac{|\omega|}{\gamma}\right)\left[ m\sqrt{m - \frac{|\omega|}{\gamma}} - \frac{1}{3}\left(m - \frac{|\omega|}{\gamma}\right)^{\frac{3}{2}} \right] \right\}. \nonumber
\end{align}

\section{Solutions of the bulk vertex's self-consistent equations} \label{app:bulk_vertex_selfcon}
Substituting Eq. \ref{eq:ansatz_bulk} into Eq. \ref{eq:vertex_bulk_eq}, we have a system of self-consistent equations for $f_i^0$ and $\vec f_i = (f_i^x,f_i^y,f_i^z)$,
\begin{align}  \label{eq:bulk_f_self-consistent}
	f^0_i\mathbbm{1} + \vec f_i \cdot \vec \sigma =& n_\mathrm{imp}u_0^2\int \frac{d^3q}{(2\pi)^3} f^0_iG^R(\omega,\vec q)\sigma_iG^A(\omega,\vec q)  \nonumber \\
	& +  n_\mathrm{imp}u_0^2\int \frac{d^3q}{(2\pi)^3} \left( f^0_iG^R(\omega,\vec q)\mathbbm{1}G^A(\omega,\vec q) + \sum_{j} f_i^jG^R(\omega,\vec q) \sigma_j G^A(\omega,\vec q) \right).
\end{align}
Using Eq. \ref{eq:Gb0}, the integrals of $G^R_b(\omega,\vec q)\mathbbm{1}G^A_b(\omega,\vec q)$ and $G^R_b(\omega,\vec q)\sigma_i G^A_b(\omega,\vec q)$ over $\vec q$ can be expanded and simplified to
\begin{align}
	\int \frac{d^3q}{(2\pi)^3} G^+_b(\omega,\vec q) \mathbbm{1} G^-_b(\omega,\vec q) &= (a_0(\omega) + a_x(\omega) + a_y(\omega) + a_z(\omega) ) \mathbbm{1}  + b(\omega)\sigma_y, \label{eq:g1g} \\
	\int \frac{d^3q}{(2\pi)^3} G^+_b(\omega,\vec q) \sigma_x G^-_b(\omega,\vec q) &= (a_0(\omega) - a_x(\omega) + a_y(\omega) - a_z(\omega) ) \sigma_x, \label{eq:gxg}  \\
	\int \frac{d^3q}{(2\pi)^3} G^+_b(\omega,\vec q) \sigma_y G^-_b(\omega,\vec q) &= (a_0(\omega) + a_x(\omega) - a_y(\omega) - a_z(\omega) ) \sigma_y + b(\omega)\mathbbm{1}, \label{eq:gyg}  \\
	\int \frac{d^3q}{(2\pi)^3} G^+_b(\omega,\vec q) \sigma_z G^-_b(\omega,\vec q) &=(a_0(\omega) - a_x(\omega) - a_y(\omega) + a_z(\omega) ) \sigma_z, \label{eq:gzg} 
\end{align}
where
\begin{align}
	&a_0(\omega) = \int \frac{\omega^2}{R(\omega,\vec q)}\frac{d^3 q}{(2\pi)^3}, \ a_x(\omega) = \int \frac{(v_z\ell(q_x))^2}{R(\omega,\vec q)}\frac{d^3 q}{(2\pi)^3}, \nonumber \\
	&a_y(\omega) = \int \frac{(v_yq_y)^2}{R(\omega,\vec q)}\frac{d^3 q}{(2\pi)^3}, \ a_z(\omega) = \int \frac{(v_zq_z)^2}{R(\omega,\vec q)}\frac{d^3 q}{(2\pi)^3}, \nonumber \\
	&b(\omega) = -2p\omega\int \frac{v_z\ell(q_x)}{R(\omega,\vec q)}\frac{d^3 q}{(2\pi)^3}. \nonumber
\end{align}
with 
\begin{align}
	R(\omega,\vec q) \equiv K(\omega,\vec q)K^*(\omega,\vec q)\nonumber
\end{align}
and
\begin{align}
K(\omega,\vec q) =&  \left(\omega + i\Gamma_{b,0}(\omega)\right)^2 - \left[(v_zq_z)^2 + (v_yq_y)^2 + (v_z\ell(q_x) + i\Gamma_{b,2}(\omega))^2\right]. \nonumber
\end{align}
In this paper, we focus on the vertex $\Lambda_x$. Substituting the integrals from Eqs. \ref{eq:g1g}, \ref{eq:gxg}, \ref{eq:gyg}, and \ref{eq:gzg}, into the system of self-consistent equations for $\vec f_i$ in the case $i = x$ (Eq.\ref{eq:bulk_f_self-consistent}), we have
\begin{align}
	f^0_x\mathbbm{1} + \vec f_x \cdot \vec \sigma &= n_{\mathrm{imp}}u_0^2(a_0 - a_x + a_y - a_z)\sigma_x + n_{\mathrm{imp}}u_0^2\bigg[(a_0 + a_x + a_y + a_z)f_x^0\mathbbm{1} +  b f^0_x \sigma_y +  (a_0 - a_x + a_y - a_z)f_x^x\sigma_x \nonumber \\
	& + (a_0 + a_x - a_y - a_z)f_x^y\sigma_y + bf_x^y\mathbbm{1}+ (a_0 - a_x- a_y + a_z)f_x^z\sigma_z \bigg] \nonumber
\end{align}
Equating the coefficients of $\mathbbm{1}$ and $\sigma_i$ on both sides, we obtain the following equations,
\begin{align}
	f^0_x &= n_\mathrm{imp}u_0^2 (a_0 + a_x + a_y+ a_z) f^0_x + n_\mathrm{imp}u_0^2 b f_x^y, \nonumber \\
	f^x_x &= n_\mathrm{imp}u_0^2 (a_0 - a_x + a_y - a_z) ( 1 + f^x_x), \nonumber \\
	f^y_x &= n_\mathrm{imp}u_0^2 (a_0 + a_x - a_y - a_z) f^y_x + n_\mathrm{imp}u_0^2 b f_x^0, \nonumber \\
	f^z_x &= n_\mathrm{imp}u_0^2 (a_0 - a_x - a_y + a_z) f^z_x, \nonumber
\end{align}
whose solutions are
\begin{align}
&f_x^0 = f_x^y = f_x^z = 0 \nonumber \\
&f_x^x = \frac{n_\mathrm{imp}u_0^2(a_0 - a_x + a_y - a_z)}{1 - n_\mathrm{imp}u_0^2(a_0 - a_x + a_y - a_z)}.
\end{align}

In order to evaluate the integrals in the $a_i$'s, we make a change of variables, $(v_z\ell(q_x),v_yq_y,v_zq_z) = |\varepsilon|\cos\theta,|\varepsilon|\sin\theta\cos\phi,|\varepsilon|\sin\theta\sin\phi)$ or $(q_{x\pm},q_y,q_z) = \left(\pm\sqrt{m - \frac{|\varepsilon|}{\gamma}\cos\theta},\frac{1}{v_y}|\varepsilon|\sin\theta\cos\phi,\frac{1}{v_z}|\varepsilon|\sin\theta\sin\phi\right)$. The $\pm$ signs refer to the two Weyl nodes which are centered at $p\vec k_0$. In order for $q_x$ to be real, one requires $\cos\theta < \frac{m\gamma}{|\varepsilon|}$. Hence, for $|\varepsilon| < m\gamma$, the angle $\theta$ is in the range, $0 < \theta <\pi$, and for $|\varepsilon| > m\gamma$, the angle $\theta$ must satisfy the inequaility, $\cos^{-1}\frac{m\gamma}{|\varepsilon|} < \theta <\pi$. We find that the Jacobian of this change of variable is
\begin{align}
	\left|\frac{\partial(q_{x\pm},q_y,q_z)}{\partial(\varepsilon,\theta,\phi)}\right| = \frac{\varepsilon^2}{2\gamma \sqrt{m}v_y v_z} \frac{\sin\theta}{\sqrt{1 - \frac{|\varepsilon|}{m\gamma}\cos\theta}}
\end{align}
The integrands in $a_i$'s are even with respect to $q_x$. This means one only needs to integrate over $q_x > 0$, i.e., $\int \frac{d^3 q}{(2\pi)^3} = 2\int_{q_x > 0} \frac{d^3 q}{(2\pi)^3}$. Furthermore, the function $R(\omega,\vec q)$ defined above is changed to
\begin{align}
	R(\omega, \varepsilon, \theta) =  K(\omega,\varepsilon,\theta)K^*(\omega,\varepsilon,\theta) \nonumber
\end{align}
with
\begin{align}
K(\omega,\varepsilon,\theta) =& \left(\omega + i\Gamma_{b,0}(\omega)\right)^2 - \left(\varepsilon^2 + 2i\varepsilon\Gamma_{b,2}(\omega)\cos\theta - \Gamma^2_{b,2}(\omega)\right). \nonumber
\end{align}
Thus, under this change of variable, the $a_i$'s transform to
\begin{align}
	a_0(\omega) &\equiv \frac{1}{8\pi^3\gamma \sqrt{m}v_yv_z} \int d(\phi,\epsilon,\theta)  \frac{\omega^2\varepsilon^2\sin\theta}{R(\omega,\varepsilon,\theta)\sqrt{1-\frac{|\varepsilon|}{m\gamma}\cos\theta}}, \label{eq:a0_lfstz_2} \\
	a_x(\omega) &\equiv \frac{1}{8\pi^3\gamma \sqrt{m}v_yv_z} \int d(\phi,\epsilon,\theta) \frac{\varepsilon^4\cos^2\theta\sin\theta}{R(\omega,\varepsilon,\theta)\sqrt{1-\frac{|\varepsilon|}{m\gamma}\cos\theta}}, \label{eq:ax_lfstz_2}  \\
	a_y(\omega) &\equiv \frac{1}{8\pi^3\gamma \sqrt{m}v_yv_z} \int d(\phi,\epsilon,\theta) \frac{\varepsilon^4\sin^3\theta\cos^2\phi}{R(\omega,\varepsilon,\theta)\sqrt{1-\frac{|\varepsilon|}{m\gamma}\cos\theta}}, \label{eq:ay_lfstz_2}  \\
	a_z(\omega) &\equiv \frac{1}{8\pi^3\gamma \sqrt{m}v_yv_z} \int d(\phi,\epsilon,\theta) \frac{\varepsilon^4\sin^3\theta\sin^2\phi}{R(\omega,\varepsilon,\theta)\sqrt{1-\frac{|\varepsilon|}{m\gamma}\cos\theta}}, \label{eq:az_lfstz_2}
\end{align}
where the integral symbol here means $\int d(\phi,\epsilon,\theta) \equiv \int_{0}^{2\pi}d\phi\int_0^{m\gamma}d\varepsilon \int_0^\pi\theta + \int_{0}^{2\pi}d\phi\int_{m\gamma}^{\infty}d\varepsilon\int_{\cos^{-1}\left(\frac{m\gamma}{\varepsilon}\right)}^{\pi}d\theta$. Performing an integral over $\phi$ in Eqs. \ref{eq:ay_lfstz_2} and \ref{eq:az_lfstz_2}, we find $a_y(\omega) = a_z(\omega)$.
\section{Scaling form} \label{app:dimless}
We can express the formula for the magnetoelectric susceptibility we obtain in Secs. \ref{sec:surface_response} and \ref{sec:bulk_response} in a scaling form by changing units of various variables to be dimensionless as 
\begin{align*}
	(k_{x}, k_y, k_z) &\rightarrow (\sqrt{m}k_{x}, \frac{m\gamma}{v_y}k_y,\frac{m\gamma}{v_z}k_z ), \\
	\omega \rightarrow m\gamma\omega, \ T &\rightarrow m\gamma T, \ \varepsilon \rightarrow m\gamma\varepsilon, \ z \rightarrow \frac{v_z}{m\gamma}z.
\end{align*}
In these new units, the dimensionless bulk eigenenergy is
\begin{align}
	\varepsilon_k &= \sqrt{k_y^2 + k_z^2 + \ell^2(k_x)}, \nonumber
\end{align}
with $\ell(k_{x}) = 1 - k_{x}^2 $. Various functions and equations relating to the surface response calculations are now written in the dimensionless fashion as
\begin{align}
	\Gamma_{STS} (k_{x}) &=  \frac{1}{\pi}\ell(k_{x})\left[\sqrt{m}-\frac{1-k_{x}^2}{\sqrt{2-k_{x}^2}}\tanh^{-1}\left(\frac{1}{\sqrt{2-k_{x}^2}}\right) \right] \nonumber \\
	\Gamma_{STB} (\omega,k_x) &= \frac{|\omega|}{8\pi^2} \int_{-1}^1 dq_{x}\int_{-\infty}^\infty d\tilde{q}_z \frac{\tilde{q}_z^2(\tilde{q}_z^2 + (\ell(q_{x})-\ell(k_{x}))^2 + \ell^2(k_{x}))}{(\tilde{q}_z^2+\ell^2(q_{x}))(\tilde{q}^2_z + \ell^2(k_{x}))}\frac{\theta(\omega^2 -(\tilde{q}_z^2 + \ell^2(q_{x})))}{\sqrt{\omega^2 -(\tilde{q}_z^2 + \ell^2(q_{x}))}} \nonumber \\
	F(\omega,z,z') &= \frac{1}{\pi}\int_{-1}^{1} ds_{x} \frac{\ell^2(s_x)e^{-2(z+z')\ell(s_{x})}}{\Gamma_s(\omega,s_{x})} \nonumber \\
	X^{(0)}(\omega,z)&=\frac{1}{4\pi^2}\int^{\sqrt{m}}_{-\sqrt{m}} dk_{x} \frac{\ell(k_{x})e^{-2\ell(k_{x})z}}{\Gamma_s(\omega,k_{x})}\ \nonumber \\
	X^{(1)}(\omega,z)&= \frac{1}{2\pi^2}\int^{\sqrt{m}}_{-\sqrt{m}} dk_{x}\int_0^\infty dz' \int_0^\infty dz_1  \frac{\ell^2(k_{x})e^{-2(z+z_1)\ell(k_{x})}}{\Gamma_s(\omega,k_{x})} f(\omega,z_1,z') \nonumber
\end{align}
Finally, the scaling form of the surface susceptibility is given by
\begin{align} \label{eq:chis_dimless}
	\chi_{s} &=  \frac{2\chi_0}{\alpha}\tilde{\chi}_s(z,\mu,T),
\end{align}
where the dimensionless function $\tilde{\chi}_s(z,\mu,T)$ is
\begin{align}
	\tilde{\chi}_{s}(z,\mu,T) &=  \int_{-\infty}^{\infty}d\omega \left(-\frac{\partial n_F(\omega-\mu)}{\partial \omega}\right) (X^{(0)}(\omega,z) + X^{(1)}(\omega,z)).
\end{align}
Here, $\chi_0 \equiv \frac{e\mu_B \sqrt{m}}{v_z}$ is the quantity with units of magnetization per electric field and $\alpha \equiv \frac{n_{\mathrm{imp}}u_0^2 \sqrt{m}}{v_y v_z}$ is the dimensionless parameter that quantifies the strength of the impurity scattering.

In the case of the bulk response, various functions are now written as
\begin{align}
	\Gamma_{b,0}(\omega) &= \frac{|\omega|}{4\pi}\left(\sqrt{1+|\omega|}-\theta(1-|\omega|)\sqrt{1-|\omega|}\right), \nonumber \\
	\Gamma_{b,2}(\omega) &= -\frac{\mathrm{sgn}(\omega)}{4\pi}\Bigg( \sqrt{1+|\omega|} - \frac{1}{3}\left(1 + |\omega|\right)^{\frac{3}{2}} - \theta\left(1 - |\omega|\right)\left( \sqrt{1 - |\omega|} - \frac{1}{3}\left(1 - |\omega|\right)^{\frac{3}{2}} \right) \Bigg), \nonumber \\
	R(\omega, \varepsilon, \theta, \alpha) &= \left[\left(\omega + i\alpha \Gamma_{b,0}(\omega)\right)^2 - \left(\varepsilon^2 + 2i \alpha\varepsilon\cos\theta\Gamma_{b,2}(\omega) - \alpha^2\Gamma_{b,2}(\omega)\right)\right] \nonumber \\ 
	& \ \ \ \ \times \left[\left(\omega -i\alpha \Gamma_{b,0}(\omega)\right)^2 - \left(\varepsilon^2 - 2i\alpha\varepsilon\cos\theta \Gamma_{b,2}(\omega)- \alpha^2\Gamma_{b,2}(\omega)\right)\right], \nonumber \\
	(a_0 - a_x )(\omega,\alpha) &=  \frac{\omega^4}{4\pi^2}\left( \int_0^{1}d\varepsilon \int_0^\pi\theta + \int_{1}^{\infty}d\varepsilon\int_{\cos^{-1}\left(\frac{1}{\varepsilon}\right)}^{\pi}d\theta \right) \frac{\sin^3\theta}{R(\omega,\varepsilon,\theta,\alpha)\sqrt{1-|\varepsilon|\cos\theta}}. \nonumber
\end{align}
The scaling form of the bulk susceptibility is 
\begin{align} \label{eq:chib_dimless}
	\chi_b = \frac{2\chi_0}{\alpha} \tilde{\chi}_b(\mu,T,\alpha),
\end{align}
where the dimensionless function $\tilde{\chi}_b(\mu,T,\alpha)$ is given by
\begin{align}
	\tilde{\chi}_b(\mu,T,\alpha) = \int_{-\infty}^{\infty}d\omega \left(-\frac{\partial n_F(\omega-\mu)}{\partial \omega}\right) \frac{1}{\pi}\frac{\alpha(a_0 - a_x)(\omega,\alpha)}{1-\alpha(a_0 - a_x)(\omega,\alpha)}.
\end{align}
\twocolumngrid

\bibliography{edelsteinweyl}
\bibliographystyle{apsrev4-2}
\end{document}